\documentclass[twocolumn,pra, aps, amsmath,amssymb,superscriptaddress]{revtex4}

\usepackage{lineno}
  \usepackage{mathptmx}
\usepackage{subfigure}
\usepackage{dcolumn}
\usepackage{amsmath,amssymb}
\usepackage{bm}
\usepackage{color}
\usepackage{latexsym}
\usepackage{epstopdf}
\usepackage{color}
\usepackage[english]{babel}
\usepackage{latexsym}
\usepackage{amsfonts}
\usepackage{psfrag,graphicx}
\usepackage{epsf}
\usepackage{subfigure}
\usepackage{amsmath}
\usepackage{amssymb}
\usepackage{amsfonts}
\usepackage{bm}
\usepackage{natbib}
\usepackage{epstopdf}

\definecolor{mygrey}{gray}{0.35}
\definecolor{myblue}{rgb}{0.2,0.2,0.8}
\definecolor{myzard}{cmyk}{0,0,0.05,0}
\definecolor{mywhite}{rgb}{1,1,1}
\definecolor{myred}{rgb}{1,0.,0.3}

\usepackage[colorlinks=true,citecolor=myblue,linkcolor=myred]{hyperref}
\def\be{\begin{equation}}
\def\ee{\end{equation}}
\def\ba{\begin{align}}
\def\enda{\end{align}}
\def\bi{\begin{itemize}}
\def\ei{\end{itemize}}

 \def\ee{\mathord{\rm e}}
 
 \def\ii{\mathord{\rm i}}

\def\half{\textstyle\frac{1}{2}}

 \def\ee{\mathord{\rm e}}
 
 \def\ii{\mathord{\rm i}}

\def\half{\textstyle\frac{1}{2}}

\renewcommand{\ii}{{\rm i}}
\renewcommand{\ee}{{\rm e}}

\def\beq{\begin{equation}}
\def\beq{\begin{equation}}
\def\eeq{\end{equation}}

 \newcommand{\ket}[1]{|#1\rangle}
 \newcommand{\bra}[1]{\langle #1|}

\begin{document}

\title{Dual trapped-ion quantum simulators: an alternative route towards exotic quantum magnets}

\author{Tobias Gra\ss}
\affiliation{ICFO - Institut de Ci\`encies Fot\`oniques, Barcelona Institute of Science and Technology, Av. C. F. Gauss 3, 08860 Castelldefels (Barcelona), Spain}

\author{Maciej Lewenstein}
\affiliation{ICFO - Institut de Ci\`encies Fot\`oniques, Barcelona Institute of Science and Technology, Av. C. F. Gauss 3, 08860 Castelldefels (Barcelona), Spain}
\affiliation{ICREA - Instituci\'{o} Catalana de Recerca i Estudis Avan\c{c}ats, Lluis Companys 23, 08010 Barcelona, Spain}

\author{Alejandro Bermudez}
\email{bermudez.carballo@gmail.com}
\affiliation{Instituto de F\'isica Fundamental, IFF-CSIC, Calle Serrano 113b, Madrid E-28006, Spain}

%\date{\today}

\begin{abstract}
  We present a route towards the quantum simulation of exotic quantum magnetism
  in ion traps by exploiting dual relations between different spin models. Our strategy allows one to start from
  Hamiltonians that can be realized with current technology, while properties of an exotic dual model are inferred
  from measurements of non-local, string-order-like, operators. The latter can be achieved from fluorescence, or from certain
  spectroscopic measurements, both of which can be combined with finite-size scaling by controlling the number of ions in the
  dual quantum simulator. We apply this concept to propose quantum simulators of frustrated quantum  magnets, and Ising
  models with multi-spin interactions. We test the validity of the idea by showing numerically that the predictions of an ideal  dual
   quantum simulator are not qualitatively modified by relevant perturbations that occur naturally in the trapped-ion scenario.
\end{abstract}
\maketitle
\begingroup
\hypersetup{linkcolor=black}
\tableofcontents
\endgroup

\maketitle

\section{Introduction}

Nowadays, the importance of the {\it Ising model} in the  theory
of statistical mechanics can  be hardly
exaggerated~\cite{Ising_rmp}.   Nonetheless, early results proved
the absence of spontaneous magnetization in the one-dimensional
(1D) model and  pointed to the lack of phase transitions also in
higher dimensions~\cite{Ising_model},  thus stimulating the
appearance of other models to explain
ferromagnetism~\cite{heisenberg_model}. The arguments against the
existence of phase transitions in higher dimensions were proven
wrong by subsequent efforts~\cite{Ising_rmp}, such as the exact
solution of the Ising model on a 2D square
lattice~\cite{ising_2d}, and eventually  turned the Ising model into  a
paradigm of critical phenomena.

Moreover, the relevance of the Ising model goes far beyond the
theory of thermal phase transitions. For instance, the partition
function of the  square-lattice  Ising model, and thus its
critical phenomena,  can be related to the ground state of the 1D
Ising model subjected to a transverse magnetic field that plays the
role of the temperature~\cite{quantum_classical_ising}. This leads
to a new kind of critical effects, i.e. {quantum phase
transitions}~\cite{sachdev_book}, where the ground state of a
system changes abruptly as a microscopic parameter  controlling
the quantum fluctuations is varied. Thanks to its exact
solution~\cite{ising_transverse_field}, this {\it quantum Ising
model} also plays a pivotal role in the theory of quantum
criticality~\cite{sachdev_book} and, despite its apparent
over-simplified appearance, has turned out to be a faithful
representation of certain magnetic materials~\cite{q_ising_exp}.

The Ising model also occupies a privileged position in the theory
of emergence in many-body systems~\cite{emergence}. In certain
geometries, such as the triangular
lattice~\cite{ising_triangular}, antiferromagnetic Ising
interactions cannot be simultaneously minimized, an effect known
as {\it frustration}. In the 3D pyrochlore lattice, this magnetic
frustration  resembles the ordering of protons in water
ice~\cite{ising_model_ice}, and inhibits the development of
magnetic order even at very low temperatures. This leads to exotic
emergent effects~\cite{frustration} that can be measured in these
so-called  spin-ice materials~\cite{spin_ice_exp},  constituting a
hallmark in the physics of  spin liquids. Moreover, if quantum
fluctuations are introduced, the wisdom is that even more exotic
properties   survive  at zero temperatures, leading to the elusive
quantum spin liquids~\cite{balents_spin_liquids}.

From this perspective, it is  thus  important to understand the
interplay of frustration and quantum fluctuations in  the Ising model.
The standard   approach  to introduce  quantum
fluctuations on the  vastly degenerate   Ising ground state considers  additional Heisenberg exchange couplings  of
tunable strength,  which flip pairs of spins and lead  to the  so-called XXZ model~\cite{fazekas_rvb}. However, this is not the unique
possibility.  For instance, one may  introduce  quantum
fluctuations by a  transverse  field leading to {\it frustrated
quantum Ising models}, which can also host quantum spin liquids in
certain lattices~\cite{2d_frustrated_qi}. As a starting point
towards this goal, envisaged in this paper, it is useful to
identify setups where the physics of simpler frustrated quantum
Ising models can be explored experimentally.

In addition to frustration, interesting many-body effects may also
arise as one departs from the standard scenario of pair-wise
interactions, and considers multi-particle couplings. For
instance, the partition function of a spin-ice Ising model that
combines  2- and 4-body interactions in a square
lattice~\cite{8vm_Ising_multi_spin} can be mapped onto  the
so-called eight-vertex model~\cite{eight_vertex_model}, which is a
paradigm for critical phenomena in ice-type models. In contrast to
the standard Ising model with pairwise
interactions~\cite{ising_2d}, this {\it Ising model with
multi-spin interactions} presents some rather exotic properties,
e.g. critical exponents  vary continuously with the strength of
the 4-body interactions, thus connecting different universality
classes. In addition to this example, we note that even more
exotic phenomena  occur when {\it (i)} quantum fluctuations or/and
{\it (ii)} higher dimensions are also considered: {\it (i)} the
combination of 4-body Ising interactions with a transverse field
in the square lattice can lead to phases with topological
order~\cite{xu_moore}. {\it (ii)} Ising models on the cubic
lattice with 4-body interactions  possess a local gauge symmetry
that forbids the use of the standard local order
parameters~\cite{wegner_ising_gauge}. By introducing quantum
fluctuations through a transverse field, such models  yield a
(3+1)-dimensional Ising Gauge
theory~\cite{fradkin_dualities_ising_lgt,kogut_lgt_rmp} through
the Euclidean generalization of the quantum-classical
mapping~\cite{quantum_classical_ising}.

From this perspective, it  is  important to understand the
interplay of multi-spin interactions with quantum fluctuations,
and its role in the emergence of exotic quantum magnets.  A useful
tool to reach this goal, again  envisaged in this paper, would be
an experimental setup where  simple  quantum Ising models with
multi-spin interactions can be implemented.

\begin{figure}
\centering
\includegraphics[width=1\columnwidth, angle=0]{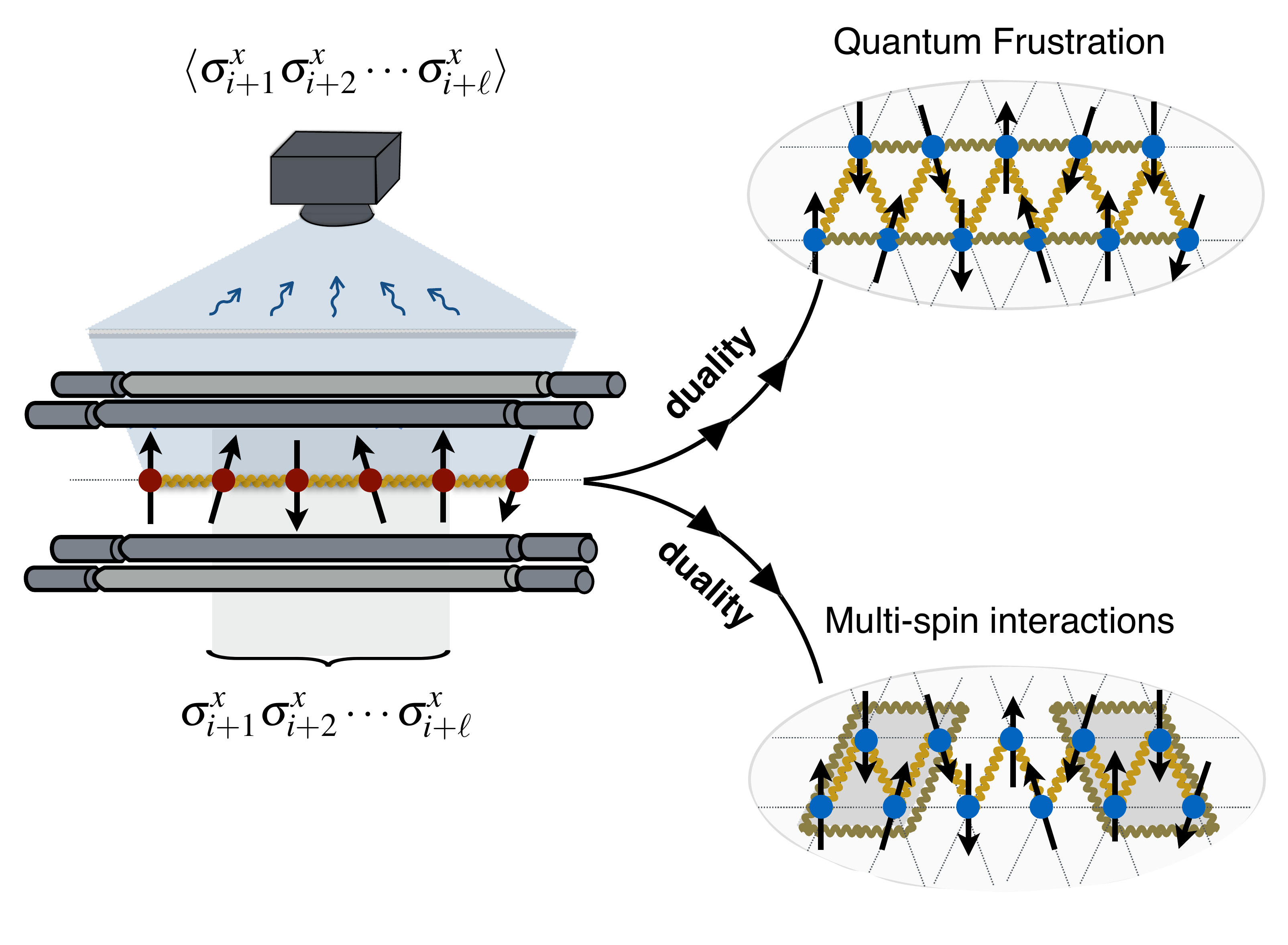}
\caption{\label{scheme_duality}
 {\bf Scheme of a trapped-ion dual quantum simulator:} The combination of phonon-mediated spin-spin interactions in linear ion chains with the ability to measure highly non-local operators would allow to exploit  a quantum duality transformation to explore exotic Ising models with tunable frustration or multi-spin interactions that can be defined in two-leg ladders with triangular motifs. }
\end{figure}

In this manuscript, we argue that current experiments of
trapped-ion crystals can address the before-mentioned goals: We
explicitly demonstrate that relevant instances  of frustrated
quantum Ising models, as well as exotic quantum Ising magnets with
multi-spin interactions, can be implemented in such systems. For
our proposal, we exploit a theoretical concept known as {\it
quantum dualities}~\cite{fradkin_dualities_ising_lgt}, which
provides a mapping between different spin models. Broadly
speaking, a dual map replaces the spin operators at the sites
(vertices) of a lattice by analog operators acting on the bonds (links)
{\it between} the vertices. Accordingly, duality can be used as a
tool to switch between local and non-local models, which, as shown
in this paper, facilitates the implementation of some Hamiltonians
of interest in a trapped-ion setup (see Fig.~\ref{scheme_duality}). 

The paper is organized as follows. In
Sec.~\ref{sec:qs_ions}, we describe two possible strategies to
simulate the required spin Hamiltonian in linear  chains of
trapped ions. In Sec.~\ref{sec:q_duality}, we review the notion of
quantum duality, highlighting its potential for quantum simulators
such as trapped-ion experiments, which allow for highly non-local
measurements. This quantum duality is applied to propose quantum
simulators of exotic quantum Ising magnets that incorporate the
effects of frustration in Sec.~\ref{sec:qs_frustration}, or
multi-spin interactions in Sec.~\ref{sec:qs_multi_spin}. Finally,
we present our conclusions and outlook in
Sec.~\ref{sec:conclusions}.

\section{Trapped-ion longitudinal XY model}
\label{sec:qs_ions}

The ever-improving experimental control of certain quantum
systems, especially in the realm of atomic, molecular, and optical
physics, has reached such a level that it is now possible to
devise systems  that will behave according to specific   quantum many-body models~\cite{feynman_qs}. This
synthetic quantum matter, usually known as a {\it quantum
simulator}, provides an alternative route to address longstanding
problems in condensed-matter physics, such as  the interplay of
frustration and quantum fluctuations mentioned above. Among the
different experimental platforms available nowadays, small
crystals of atomic ions confined by electromagnetic
traps~\cite{wineland_review} have already proven to be quite
flexible  quantum simulators~\cite{qsim_ions}. Let us briefly
review the possibilities of  trapped-ion quantum simulators in the
context of frustrated and multi-spin quantum Ising models.

Following the seminal proposal to use the phonons of the ion
crystal as mediators of  spin-spin
interactions~\cite{spin_models_porras} (for earlier related
proposals see \cite{spin_models_wunderlich}), small-scale   1D
quantum Ising magnets have been experimentally
realized~\cite{ising_exps}. The most direct approaches to include
frustration would require  engineering 2D ion lattices with
triangular motifs, either by considering Penning
traps~\cite{spin_models_porras,frustrated_ising_penning} and
micro-fabricated trap arrays in a surface
electrode~\cite{spin_models_porras,surface_traps}, or by coping
with the excess micro-motion associated with 2D lattices in common
rf-traps~\cite{ising_ladders_proposal}. A different approach,
which has proven very
fruitful~\cite{frustrated_ion_chain_long_range},   retains the 1D
geometry of the crystals in linear Paul traps while  exploiting
the long-range character of the Ising interactions to induce
frustration via  next-to-nearest-neighbor couplings. In this work,
we shall introduce an alternative method  that also permits
retaining the 1D  lattices, but offers more flexibility by
allowing full control of the range of frustration in the simulated
spin models.

Regarding multi-spin interactions, there have been
proposals~\cite{3_spin_ions} that generalize the above
schemes~\cite{spin_models_porras} by considering non-linear
spin-phonon couplings that can induce 2- and 3-body Ising
interactions at the expense of getting  weaker spin-spin couplings
with respect to the pair-wise scenario. Hence, synthesizing other
multi-spin interactions, such as the mentioned 4-body Ising
couplings, will result in even weaker coupling strengths that
question the possibility of an experimental observation. A
different and very fruitful approach is to build  stroboscopic
quantum simulators by concatenating well-controlled
unitaries~\cite{stroboscopic_qs}. For the particular unitaries
available in trapped-ion setups, multi-spin interactions can be
obtained by introducing simple sequences that involve auxiliary
spins~\cite{multi_spin_dissipative}, as demonstrated in  the
experiments~\cite{multi_spin_dissipative,stroboscopic_multi_spin_ions}.
In this work, we propose an alternative method  to implement
4-body interactions  in an analog fashion,  thus not limited by
the accumulated error of the concatenated unitaries, but without
the above-mentioned  limitation on the coupling strengths.

The central idea of this work is to exploit the concept of quantum
duality~\cite{fradkin_dualities_ising_lgt} to reach the desired
quantum simulator of exotic  Ising models with frustration, or
multi-spin interactions, starting from a different Hamiltonian
that is easier to realize in a certain setup. In particular, we
will focus on the following physical Hamiltonian \beq
\label{dipolar_longitudinal_XY} H_{\rm \ell
XY}=\sum_{i<j}\left(J_{ij}^x\sigma_i^x\sigma_j^x+J_{ij}^y\sigma_i^y\sigma_j^y\right)-\sum_ih\sigma_i^x,
\eeq which shall be referred to as the {\it longitudinal XY model}
($\ell$XY), and is the one to be implemented in the ion-trap
setup. Let us remark on the two unique features that make  this
synthetic magnet  very different from real magnetic materials:
{\it (i)} the exact number of spins can be experimentally
controlled, {\it (ii)} the magnetization, or any other correlation
function, can be measured locally. These two features, in
combination with the  quantum duality, will be crucial for the
quantum simulation scheme hereby proposed. 

Let us now describe the
required ingredients to realize the above
Hamiltonian~\eqref{dipolar_longitudinal_XY}, and postpone the
duality connection to the desired Ising models to the following
sections. We start by describing the standard approach in
Sec.~\ref{two_forces}, and then introduce an alternative that
might be advantageous in Sec.~\ref{driving}.

\subsection{Spin model from a pair of spin-dependent forces}
\label{two_forces}

 We consider a collection of laser-cooled atomic ions confined  by a linear Paul trap, and forming
 a 1D Coulomb crystal with equilibrium positions ${\bf r}_i^0=z_i^0{\bf e}_z$ for $i\in\{1,\cdots,N\}$~\cite{james_ion_crystal} (see Fig.~\ref{scheme_duality}).
 The small vibrations around  these equilibrium positions can be described in terms of three phonon branches, two radial
 $\alpha\in\{x,y\}$ and one longitudinal $\alpha=z$, namely
\beq \label{eq:Ho_phonons} H_{\rm
p}=\sum_{n,\alpha}\omega_{n,\alpha}a_{n,\alpha}^{\dagger}a_{n,\alpha}^{\phantom{\dagger}},
\eeq where we have introduced the bosonic  operators
$a_{n,\alpha}^{\dagger}, a_{n,\alpha}^{\phantom{\dagger}}$ that
create-annihilate a phonon associated to the normal-mode frequency
$\omega_{n,\alpha}$, where $n\in\{1,\cdots,N\}$.

The spin chain is formed by  a pair of  long-lived electronic
levels  from the energy-level structure of each particular ion,
denoted as $\ket{{\uparrow}_i},\ket{{\downarrow}_i}$. This leads
to a trivial spin Hamiltonian \beq \label{eq:Ho_spins} H_{\rm
s}=\sum_i\frac{\omega_0}{2}\sigma_i^z, \eeq where we have
introduced the transition frequency $\omega_0$, and the Pauli spin
operator
$\sigma_i^z=\ket{{\uparrow}_i}\bra{{\uparrow}_i}-\ket{{\downarrow}_i}\bra{{\downarrow}_i}$.
These spins can be  coupled  by phonon-mediated interactions,
which require a particular form of spin-phonon coupling. One
typically considers the so-called spin-dependent dipole forces,
which follow from a moving optical lattice, and read as follows
\beq \label{spin_dep_forces} H_{\rm
sp}=\sum_{i,n}\sum_{\alpha,\beta} F_{i,n}^{\alpha,\beta}\Delta
r_{n,\alpha}\sigma_i^\beta
a_{n,\alpha}^{\dagger}\ee^{\ii\Delta_{n,\alpha,\beta}t}+{\rm
H.c.}, \eeq where we work in the interaction picture with respect
to Eqs.~\eqref{eq:Ho_phonons}-\eqref{eq:Ho_spins}, and have
neglected rapidly oscillating terms within a rotating wave
approximation. Here, we have introduced the two remaining Pauli
spin operators
$\sigma_i^x=\ket{{\uparrow}_i}\bra{{\downarrow}_i}+{\rm H.c.,}$
$\sigma_i^y=\ii\ket{{\downarrow}_i}\bra{{\uparrow}_i}+{\rm H.c.,}$
the zero-point motion of the different vibrational modes $\Delta
r_{n,\alpha}$,  the dipole light forces $F_{i,n}^{\alpha,\beta}$,
and their corresponding detunings $\Delta_{n,\alpha,\beta}$. We
shall assume that a particular light force will only couple to a
single vibrational axis,  namely
$F_{i,n}^{\alpha,\beta}=F_{i,n}^{\alpha}\delta_{\alpha,\beta}$.

If the trap frequencies are different
$\omega_x\neq\omega_y\neq\omega_z$, such that their differences
are much bigger than the far-detuned dipole forces
$|F_{i,n}^{\alpha,\beta}\Delta r_{n,\alpha}|\ll
|\Delta_{n,\alpha,\beta}|\ll |
\omega_{\alpha}-\omega_\beta|$~\cite{spin_models_porras}, the
phonons act as mere carriers of spin-spin  interactions \beq
\label{eff_ham} H_{\rm eff}=\sum_{i<
j}\left(J_{ij}^x\sigma_i^x\sigma_j^x+J_{ij}^y\sigma_i^y\sigma_j^y+J_{ij}^z\sigma_i^z\sigma_j^z\right),
\eeq where the phonon-mediated spin-spin couplings \beq
\label{couplings} J_{ij}^{\alpha}=-\sum_{n}\frac{F_{i,n}^{\alpha}
(F_{j,n}^{\alpha})^*\Delta r_{n,\alpha}^2}{\Delta_{n,\alpha}}+{\rm
c.c.}, \eeq have the standard second-order scaling
$\mathcal{O}(F^2/\Delta)$, and thus correspond to virtual
processes where a phonon is created and subsequently  absorbed by
a pair of distant spins in the chain.

We are interested in anisotropic XY
models~\eqref{dipolar_longitudinal_XY}, and thus shall only
consider dipole forces that couple to the two radial branches
$F_{i,n}^{x}$, and $F_{i,n}^{y}$, such that the effective
Hamiltonian~\eqref{eff_ham} only contains XX and YY couplings. In
addition, we shall consider a so-called longitudinal field $h$,
which can be implemented by considering the standard carrier
transitions~\cite{wineland_review} with the correct phase. The
combination of these ingredients yields the desired $\ell$XY model
of Eq.~\eqref{dipolar_longitudinal_XY} with the corresponding
spin-spin couplings in Eq.~\eqref{couplings}. As announced above,
we shall not exploit the long-range nature of such couplings to
engineer frustration~\cite{frustrated_ion_chain_long_range}.
Instead, we will consider the largest possible detunings, which
lead to the smallest possible errors in the quantum
simulation~\cite{spin_models_porras}, and also yield a
fast-decaying dipolar tail of the anti-ferromagnetic spin-spin
interactions \beq \label{dip_decay} J_{ij}^{\alpha}\approx
\frac{J_0^\alpha}{|z_i^0-z_j^0|^3},\hspace{2ex}
J_0^\alpha>0,\hspace{2ex}\alpha\in\{x,y\}. \eeq We thus obtain a
dipolar version of the famous XY model~\cite{xy_model}, subjected
to an additional longitudinal field.

\subsection{Spin model from a single spin-dependent force }
\label{driving}

Let us note that, so far, there has not been any experimental
realization of a  spin model exploiting simultaneously two phonon
branches of a trapped-ion crystal. Instead, the dynamics of
excitations in the isotropic limit of the XY model has been
explored by considering a single Ising model, and thus a single
phonon branch to mediate the interactions, subjected to an
additional very strong transverse
field~\cite{xy_model_ions,xy_ions_spect}. Unfortunately, this
isotropic limit  leads to $J_{ij}^x=J_{ij}^y$ in the above
Hamiltonian~\eqref{dipolar_longitudinal_XY}. As will become
apparent in the following sections, this would limit its interest
as a simulator of exotic quantum magnets under the duality
transformation. We now introduce a scheme to modify the
implementation of Refs.~\cite{xy_model_ions,xy_ions_spect} and
obtain a fully-tunable XY model in a longitudinal field by
exploiting a single phonon branch.

Let us consider the Ising model in a longitudinal field, which
arises from considering a spin-dependent dipole force coupled to a
single radial branch and a carrier transition \beq
\label{long_ising} H_{\rm \ell I}=\sum_{i<
j}\tilde{J}_{ij}^x\sigma_i^x\sigma_j^x-\sum_i\tilde{h}\sigma_i^x.
\eeq Instead of applying a strong transverse field to obtain the
isotropic XY model~\cite{xy_model_ions,xy_ions_spect}, let us
consider a periodically-modulated transverse field, which may
arise from the cross-beam ac-Stark shift from a pair of
co-propagating lasers with different frequencies. This is
described by \beq H_{\rm m}(t)=\sum_{i}\frac{\Omega_{\rm
d}}{2}\cos(\omega_{\rm d}t)\sigma_i^z, \eeq where $\Omega_{\rm d}$
is the strength of the transverse field, and $\omega_{\rm d}$ its
modulation frequency, although we note that the following argument
could as well be applied for a transverse detuned carrier by
simply exchanging  $\sigma_i^z\to\sigma_i^y$.

We start by moving    to an interaction picture with respect to
the driving, to obtain the following dressed Hamiltonian \beq
H_{\ell
I}(t)=\sum_{i<j}\tilde{J}_{ij}^x\sigma_i^+\sigma_j^-+\sum_{i<j}\tilde{J}_{ij}^x\frak{f}(t)\sigma_i^+\sigma_j^+-\sum_i
\tilde{h}\sqrt{\frak{f}(t)}\sigma_i^++{\rm H.c.}, \eeq where we
have introduced the modulation function  $\frak{f}(t)={\rm
exp}\{\ii\frac{\Omega_{\rm d}}{\omega_{\rm d}}\sin(\omega_{\rm
d}t)\}$, and the  operators
$\sigma_i^{\pm}=\half(\sigma_i^x\pm\ii\sigma_i^y)$. By using the
Jacobi-Anger expansion of the exponential in terms of Bessel
functions, this modulation function can be expressed as
$\frak{f}(t)=\sum_{n\in\mathbb{Z}}\frak{J}_{n}(\frac{\Omega_{\rm
d}}{\omega_{\rm d}})\ee^{\ii n\omega_{\rm d}t}$. We can now
simplify the Hamiltonian considerably if we consider that the
modulation frequency is much larger than the couplings of the
Ising model~\eqref{long_ising}, namely $|\tilde h|,|\tilde
J_{ij}^x|\ll\omega_{\rm d}$. By applying a rotating-wave
approximation, we find a time-independent Hamiltonian \beq
\begin{split}
H_{\ell I}(t)&=\sum_{i<j}J_{ij}^x\sigma_i^+\sigma_j^-+\sum_{i<j}J_{ij}^x\frak{J}_{0}\left(\Omega_{\rm d}/\omega_{\rm d}\right)\sigma_i^+\sigma_j^+\\
&-\sum_i h\frak{J}_{0}\left(\Omega_{\rm d}/2\omega_{\rm d}\right)\sigma_i^++{\rm H.c.},
\end{split}
\eeq where one observes that some of the coupling strengths are
dressed by the zero Bessel function. By moving back to the
standard Pauli spin operators, we obtain the desired $\ell$XY
model~\eqref{dipolar_longitudinal_XY}, $H_{\ell I}(t)=H_{\rm \ell
XY}$, with the following parameters \beq
\begin{split}
J_{ij}^x&=\tilde{J}_{ij}^x\big(1+\frak{J}_{0}\left(\Omega_{\rm d}/\omega_{\rm d}\right)\big)/2,\\
J_{ij}^y&=\tilde{J}_{ij}^x\big(1-\frak{J}_{0}\left(\Omega_{\rm d}/\omega_{\rm d}\right)\big)/2,\\
h_{\phantom{ij}}&=\tilde{h}\frak{J}_{0}\left(\Omega_{\rm d}/2\omega_{\rm d}\right).
\end{split}
\eeq

Hence, by controlling two ratios $\tilde{h}/\tilde{J}_0^x$, and
$\Omega_{\rm d}/\omega_{\rm d}$, we can access the full phase
diagram of the target  $\ell$XY
model~\eqref{dipolar_longitudinal_XY}. We note that this scheme
might be considered as a  spin version of the so-called coherent
destruction of tunnelling for electrons in solids under periodic
electric fields~\cite{cdt}, and is the simplest possible
modulation scheme that can lead to interesting effective
Hamiltonians. In the context of trapped ions, other  driving terms
have been exploited to simulate the effects of synthetic gauge
fields in the vibrational or spin
sectors~\cite{periodic_modulations_phonons,periodi_modulations_spins}.

Before moving to the next section, where we  exploit the tool of
quantum dualities for quantum simulations, let us summarize the
ingredients that we have introduced so far. {\it (i)} By loading
the ion crystal in a controlled fashion, we can design the number
of spins $N$ in the synthetic quantum magnets as desired. {\it
(ii)} By combining the resonance fluorescence in a cycling
transition~\cite{wineland_review} with global single-spin
rotations in the Bloch sphere (i.e. single-qubit gates), we can
measure the observables $\langle \sigma_i^\alpha \rangle$,
$\langle\sigma_{i}^{\alpha}\sigma_{j}^\alpha\rangle$,
$\langle\sigma_{i}^{\alpha}\sigma_{j}^\alpha\sigma_{k}^\alpha\rangle,
\langle\sigma_{i}^{\alpha}\sigma_{j}^\alpha\sigma_{k}^\alpha\sigma_{l}^\alpha\rangle$,
and so on. {\it (iii)} By using carrier transitions and
far-detuned spin-dependent dipole forces, either along two
different vibrational axes with different trap frequencies, or
combining a single force with a periodic modulation, we can obtain
a synthetic longitudinal XY model with dipolar
couplings~\eqref{dipolar_longitudinal_XY}. These three ingredients
will be fundamental to propose a dual quantum simulator in the
following section. We shall argue that, even if the dipolar  XY
model in a longitudinal field does not look  more exotic than the
standard quantum Ising model, it can lead to very interesting
phenomena such as quantum frustration when complemented by quantum
dualities and certain local measurements which, although
complicated for other setups, are accessible in trapped-ion
experiments.

\section{Duality for quantum simulations}
\label{sec:q_duality}

The notion of duality in physics, which first appeared in
connection to the equations of electromagnetism in the absence of
sources, has become a far reaching concept in different
disciplines. In this work, we shall focus on the use of dualities
to understand the properties of many-body lattice models, which
started with a seminal work on the self-duality of the  Ising
model  on the square lattice~\cite{classical_duality_ising_2d}.
This property  permitted locating the critical temperature of the
model prior to its exact solution~\cite{ising_2d}, and turned out
to be a powerful concept that can be generalized to other
situations~\cite{classical_dualities_review}, including  also  the
quantum-mechanical
regime~\cite{fradkin_dualities_ising_lgt,kogut_lgt_rmp}.

\subsection{Quantum duality transformation}

In essence, a duality transformation relates the properties of a
particular model in the original lattice $\Lambda$ with those of a
transformed model  in the {\it dual lattice} $\Lambda^*$. In 1D,
which is the regime of interest to our work,  the dual lattice is
obtained by placing vertices at the links of the original lattice,
and setting bonds between those vertices separated by a site of
the original lattice. Therefore, the dual lattice of a 1D chain is
simply another chain that has been translated by half a lattice
constant. Typically, this transformation is defined for infinite
Bravais lattices, and aims at giving a different perspective of
the model under study in the so-called thermodynamic limit.
However, since we are interested in small-scale synthetic magnets,
one needs to consider the effects of boundary conditions, which
map an original $N$-site chain $\Lambda_{N}$ onto an $(N+1)$-site
dual chain $\Lambda^*_{N+1}$ \beq \label{dual_lattice}
\begin{split}
\Lambda_N&:\put(9,2.5){{\line(1,0){80}}}\hspace{1.65ex}\bullet\hspace{2ex}\bullet\hspace{2ex}\bullet\hspace{2ex}\bullet\hspace{2ex}\bullet\hspace{2ex}\bullet\hspace{0.2ex}\cdots\hspace{0.2ex}\bullet\put(-2,2.5){{\line(1,0){12}}}\hspace{2ex}\bullet\\
\phantom{\Lambda^*_{N+1}}&\phantom{:\put(9,2.5){{\line(1,0){80}}}}\hspace{1.65ex}\hspace{0.75ex}\textcolor{mygrey}{\downarrow}\hspace{0.2ex}\hspace{2ex}\hspace{0.2ex}\textcolor{mygrey}{\downarrow}\hspace{0.2ex}\hspace{2ex}\hspace{0.2ex}\textcolor{mygrey}{\downarrow}\hspace{0.2ex}\hspace{2ex}\hspace{0.1ex}\textcolor{mygrey}{\downarrow}\hspace{0.1ex}\hspace{2ex}\hspace{0.1ex}\textcolor{mygrey}{\downarrow}\hspace{0.1ex}\hspace{2ex}\hspace{0.1ex}\textcolor{mygrey}{\downarrow}\hspace{0.1ex}\phantom{\cdots}\hspace{4.5ex}\textcolor{mygrey}{\downarrow}\\
\Lambda^*_{N+1}&:\hspace{2ex}\put(2,2.5){{\line(1,0){95}}}\bullet\hspace{2ex}\bullet\hspace{2ex}\bullet\hspace{2ex}\bullet\hspace{2ex}\bullet\hspace{2ex}\bullet\hspace{2ex}\bullet\hspace{0.2ex}\cdots\hspace{0.2ex}\bullet\put(-2,2.5){{\line(1,0){12}}}\hspace{2ex}\bullet\\
\end{split}
\eeq Once these lattices have been defined, let us introduce the
quantum duality  on the spin
operators~\cite{fradkin_dualities_ising_lgt,kogut_lgt_rmp}. The
$\frak{su}(2)$ algebra of the spin operators is preserved by
defining the dual spin operators as follows \beq
\label{dual_operators} \tau_i^z=\prod_{j\leq
i}\sigma_j^x,\hspace{2ex}
\tau_i^x=\sigma_{i}^y\sigma_{i+1}^y,\hspace{2ex}\forall
i\in\Delta^*_{N+1}. \eeq Then, the dual Hamiltonian to the
$\ell$XY model~\eqref{dipolar_longitudinal_XY} becomes \beq
\label{dual_lXY} H^{\rm dual}_{\rm \ell
XY}=\sum_{i>j}\left(J_{ij}^x\tau_{i-1}^z\tau_{i}^z\tau_{j-1}^z\tau_{j}^z+J_{ij}^y\prod_{i\leq
k< j}\tau_k^x\right)-\sum_ih\tau_{i-1}^z\tau_{i}^z, \eeq where the
sums must be extended to the dual lattice $\Lambda^*_{N+1}$, and
thus to $N+1$ sites. At first sight, this dual mapping does not
seem to produce any interesting result, since the dual Hamiltonian
seems to be more convoluted than the original one. However,
exploiting the fast dipolar decay and some additional properties
of the spin-spin couplings~\eqref{dip_decay}, this dual
Hamiltonian can host a variety of exotic effects. Let us postpone
this discussion for the following sections, and address first a
question of relevance for the dual quantum simulator.

\subsection{Quantum dual measurements}
 Usually, quantum dualities are used as theoretical
 tools that allow to understand certain aspects of a particular lattice model from a different perspective.
 In this manuscript, we are instead proposing to use them  experimentally, which is usually hampered by the non-local
 character of the dual operators~\eqref{dual_operators}. In order to understand the properties of the dual model~\eqref{dual_lXY},
 one typically  studies  two-point correlators in the dual basis, which become highly-nonlocal correlators in the original lattice, such as
\beq \label{corr_functions}
\langle\tau_i^z\tau_{i+\ell}^z\rangle=\langle
\sigma_{i+1}^x\sigma_{i+2}^x\cdots\sigma_{i+\ell}^x\rangle. \eeq
Therefore, in order to exploit the quantum duality experimentally
and understand properties of the model~\eqref{dual_lXY} by
performing experiments with the physical
Hamiltonian~\eqref{dipolar_longitudinal_XY}, one needs to measure
{\it string-order-like parameters}, which is usually impossible in
most physical setups. However, as argued in the previous sections,
these type of observables are exactly the ones that can be
obtained via the combination of single-qubit gates and
fluorescence in trapped-ion experiments.

Apart from correlation functions, also spectral properties give
interesting insight into the behavior of a physical system, and
important properties such as energy gaps of low-lying excitations
are the same for the original model and its dual. Accordingly,
spectroscopic measurements in the spirit of the recent
experiments~\cite{xy_ions_spect,spectroscopy_many_body} may
provide an alternative way of obtaining relevant information valid
for both models.

Before entering into the discussion about the exotic magnetic
phases that the Hamiltonian~\eqref{dual_lXY} can host, let us also
comment on the other highly-unusual property of our synthetic
quantum magnets: the possibility of exactly controlling the number
of spins $N$.  Critical phenomena and quantum phase transitions
take place in the  thermodynamic limit ($N\to \infty$). Although
this is a mathematical idealization since real systems are always
finite, this type of predictions  agree considerably well with
those based on moderately-large  samples, as predicted by {\it
finite size scaling} (FSS)~\cite{fss}. Although, strictly
speaking,  phase transitions can only  occur  at the thermodynamic
limit, FSS  can be considered as  a window to extract  their
infinite-size characteristics (e.g. critical points and exponents)
by studying the scaling of certain observables with the system
size. As occurs with quantum dualities, FSS  has been mainly a
theoretical tool used in combination with numerical simulations of
small  systems of increasing  size. On the other hand,
experimental FSS in real materials would be hampered by
difficulties on {\it (i)} controlling accurately $N$, and {\it
(ii)} distinguishing  finite-size effects from other spurious
rounding effects (e.g. disorder, impurities, etc.). This situation
has recently changed with  trapped-ion
experiments~\cite{ising_exps}, where the number of Ising spins can
be controlled exactly, and where  typical disorder and  impurity
effects are totally absent. Moreover, the possible rounding
effects caused by other sources of noise can be experimentally
controlled and identified.

In the following two sections, we will combine the duality
transformation, FSS, and non-local observables, to show that the
mild-looking Hamiltonian~\eqref{dipolar_longitudinal_XY} can
indeed simulate a variety of exotic magnetic phenomena.

\section{Dual quantum simulators of frustration}
\label{sec:qs_frustration}

In this section, we want to show that the
dual quantum simulator
can be used to explore  the interplay of frustration and quantum
fluctuations. Let us first start with a plausibility argument. One
can rearrange  the dual Hamiltonian~\eqref{dual_lXY} as \beq
H^{\rm dual}_{\rm \ell XY}=H_{\rm
qANNNI}(-h,J_{i,i+1}^x,J_{i,i+1}^y)+\Delta H, \eeq where $H_{\rm
qANNNI}(-h,J_{i,i+1}^x,J_{i,i+1}^y)$ is the so-called {\it quantum
axial next-nearest-neighbor Ising Hamiltonian} (qANNNI), to be
defined below. If the $\ell
$XY-model~\eqref{dipolar_longitudinal_XY} was restricted to
nearest-neighbour interactions, i. e. if $J_{ij}^\alpha \propto
\delta_{\langle i,j \rangle}$, the dual mapping between $\ell$XY
and qANNNI model would be perfect, that is, the remainder $\Delta
H=0$ would vanish. However, since our starting point is the
$\ell$XY model with dipolar interactions, $\Delta H$ will be
non-zero introducing a perturbation to the dual mapping.

The Hamiltonian of the qANNNI model is the one of a transverse
Ising model with frustrated nearest-neighbor and
next-nearest-neighbor interactions \beq \label{qANNNI} H_{\rm
qANNNI}(J_1,J_2,B)=\sum_i\left(J_1\tau_i^z\tau_{i+1}^z+J_2\tau_i^z\tau_{i+2}^z+B\tau_i^x\right).
\eeq Here, $J_1<0$ is a ferromagnetic nearest-neighbor coupling
that corresponds to the longitudinal field  of the original
Hamiltonian~\eqref{dipolar_longitudinal_XY}. This ferromagnetic
interaction   competes against an antiferromagnetic $J_2>0$
next-nearest-neighbor coupling, corresponding to the
nearest-neighbor XX couplings of the original model, and a
transverse magnetic field $B$, corresponding to the
nearest-neighbor YY couplings of the original model.  Note that these competing Ising interactions can be described in a two-leg ladder with triangular motifs, which clarifies the origin of the frustration (see Fig.~\ref{scheme_duality}). The
parameter equivalence under the duality  is thus \beq
\label{dual_param} J_1=-h,\hspace{2ex}
J_2=J_{i,i+1}^x,\hspace{2ex}B=J_{i,i+1}^y. \eeq

 Let us emphasize that the frustration range of the qANNNI model, given by the ratio of next-nearest- to nearest-neighbor
 couplings $J_2/|J_1|$, corresponds to the ratio of two different couplings $J_{i,i+1}^x/h$ in the original model~\eqref{dipolar_longitudinal_XY}
 that  arise from totally different sources, i.e. a spin-dependent force and a carrier transition. Accordingly, it is  simpler to reach all
 regimes of physical interest $J_2/|J_1|\lessgtr 1/2$ by tuning this ratio in the physical model~\eqref{dipolar_longitudinal_XY}, than it
 would be to reach them by shaping the long-range tail of a bare quantum Ising model~\cite{frustrated_ion_chain_long_range}. In fact,
 as $J_2/|J_1|\to1$, it would be no longer justified to neglect the remaining longer-range tail of the bare Ising model, as these additional
 terms  are likely to modify severely the phase diagram. Our method of controlling the range of frustration also seems more straightforward
 than modifying the crystalline structure in Penning~\cite{frustrated_ising_penning} and surface~\cite{surface_traps} traps, or changing
 the direction of the lasers/tilting the  crystal~\cite{ising_ladders_proposal} in  common rf-traps.

In addition to the qANNNI Hamiltonian, the long-range nature of the
interactions in~\eqref{dipolar_longitudinal_XY} leads to the
following perturbation \beq \Delta
H=\sum_{i=1}^{N+1}\sum_{j=i+2}^{N+1}\left(J_{ij}^x\tau_{i-1}^z\tau_{i}^z\tau_{j-1}^z\tau_{j}^z+J_{ij}^y\prod_{i\leq
k< j}\tau_k^x\right), \eeq which contains the dipolar tail of the
phonon-mediated interactions starting form the
next-nearest-neighbor couplings. Due to the fast dipolar
decay~\eqref{dip_decay}, the next-nearest-neighbor couplings are
almost one order of magnitude smaller than the nearest-neighbor
terms in Eq.~\eqref{qANNNI}, and the longer-range terms get
reduced even further. Hence, one can argue that this perturbation
will not modify in any essential manner the phase diagram of the
qANNNI model~\eqref{qANNNI}. Although the duality between the
nearest-neighbor models is already known to bear accurate results,
we will show below by numerical calculation that our plausibility
argument is correct, and that the additional long-range
contributions do not modify the usefulness of the duality mapping
for quantum simulation purposes.

The model in Eq.~\eqref{qANNNI} is the quantum-mechanical version
of the classical ANNNI model~\cite{annni_classical,annni_review}
in a square lattice, as can be proved by applying the
quantum-classical mapping~\cite{quantum_classical_ising} in the
present case~\cite{annni_classical_quantum}. The classical 2D
model, which describes stacked Ising chains with competing
ferromagnetic nearest-neighbor and anti-ferromagnetic
next-nearest-neighbor couplings at finite temperatures, is
considered to be a paradigm in the physics of magnetic frustration
and commensurate/incommensurate  phases (i.e. spatially-modulated
arrangements of the magnetic dipoles with a period that is
commensurate/incommensurate with the
lattice)~\cite{commensurate_incomm_annni_review}. In addition to
the typical second-order phase transitions describing
order-disorder phenomena in Ising magnets, this model  displays
different lines separating the commensurate/incommensurate phases,
such as Kosterlitz-Thouless~\cite{kt} and
Pokrovsky-Talapov~\cite{pt} phase transitions, or a disorder
line~\cite{peschel_emery_line} connecting unmodulated/modulated
disordered paramagnets. All these thermal phenomena find an
analogue in the ground state of the  quantum-mechanical
model~\eqref{qANNNI}, where the transverse field $B$ plays the
role of the temperature by introducing quantum fluctuations that
drive the different phase transitions.

In the following, we make an exhaustive numerical study to assess
the importance of the dipolar terms in the trapped-ion
realization, and the possibility of observing such a rich phase
diagram with state-of-the-art experimental conditions.

\begin{figure}
\centering
\includegraphics[width=0.75\columnwidth, angle=0]{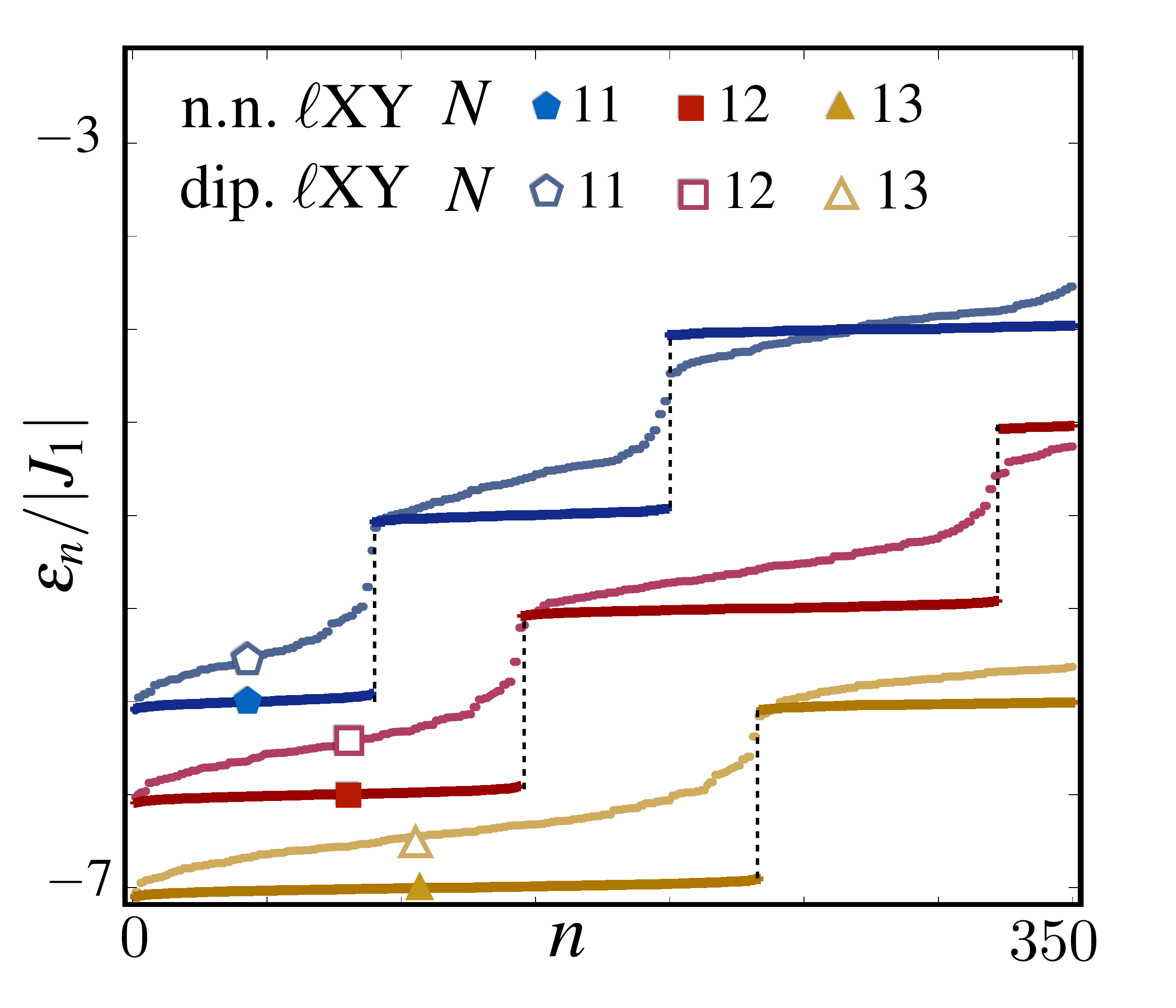}
\caption{\label{frust}
 {\bf Quasi-degenerate energy manifolds:} We plot the energies of the lowest 350 eigenstates of the longitudinal XY  model at the frustration point $J_2/|J_1|=1/2$ and $B=0$ for dipolar interactions and for nearest-neighbor interactions for system sizes between 11 and 13.}
\end{figure}

\subsection{Large ground state degeneracy}

A hallmark of frustration, sometimes even regarded as its defining
property, is the existence of a vastly-degenerate ground state
manifold that a arises from the impossibility of minimizing
simultaneously the conflicting interactions~\cite{frustration}. In
the dual model~\eqref{qANNNI}, this frustration arises from the
competition between  ferromagnetic nearest-neighbor and
anti-ferromagnetic next-nearest-neighbor interactions. Triads of
sites $(i,i+1,i+2)$ define a triangular motif, where the Ising
interactions for $J_2/|J_1|=1/2$ at $B=0$ cannot be simultaneously
minimized, and thus become frustrated. In fact, the number $D_{\rm
qANNNI}(N)$ of degenerate ground states at this frustration
point~\cite{degeneracy_annni} grows exponentially with system size
$N$ \beq \label{deg_dual} D_{\rm
qANNNI}(N)=\frac{2}{\sqrt{5}}\left[ \left( \frac{1+\sqrt{5}}{2}
\right)^{N} + \left( \frac{1-\sqrt{5}}{2} \right)^{N} \right].
\eeq

According to the duality mapping, the nearest-neighbor limit of
the XY model in a longitudinal
field~\eqref{dipolar_longitudinal_XY} should show the same
exponential growth in the ground state manifold for $h=0$, and
$J^x_{i,i+1}=2J^y_{i,i+1}$. Note, however, that the duality
transformation~\eqref{dual_operators} removes the
$\mathcal{Z}_2$-symmetry of the Ising model, and thus halves the
number of degenerate states in the original model \beq \label{deg}
D_{\ell {\rm XY}}(N)= \frac{D_{\rm qANNNI}(N)}{2}. \eeq As
discussed above, for open boundary conditions, one also needs to
take into account the fact that the duality maps a system of $N$
spins onto a system with $N+1$ bond operators. We also note that
it is possible to preserve the $\mathcal{Z}_2$ symmetry by
modifying the duality transformation such that the transverse
field is not applied to all spins~\cite{bond_dualities}, but here
we stick to the standard duality
transformation~\eqref{dual_operators}.

We have verified the counting of Eq. (\ref{deg}) for small systems
up to $N=14$, where the XY interaction is truncated to nearest
neighbors. Turning our attention to the  dipolar $\ell$XY
model~\eqref{dipolar_longitudinal_XY}, the addition of the
long-range tail is especially disturbing around this
maximally-frustrated regime, as the vast number of degenerate
ground states turns any perturbation, no matter how small, into a
highly non-perturbative problem. Nevertheless, the dipolar system
still exhibits some traces of the exponentially large degeneracies
in the ``clean'' model. As demonstrated in  Fig.~\ref{frust} by
contrasting the 350 lowest eigen-energies of the dipolar model to
the energies of the nearest-neighbor model, the dipolar system
exhibits a low-energy manifold separated from higher states by a
spectral gap, and the dimension of this manifold is precisely
given by Eq. (\ref{deg}).

\subsection{Phase diagram for finite chains}

\begin{figure*}[t]
\centering
\includegraphics[width=1\textwidth, angle=0]{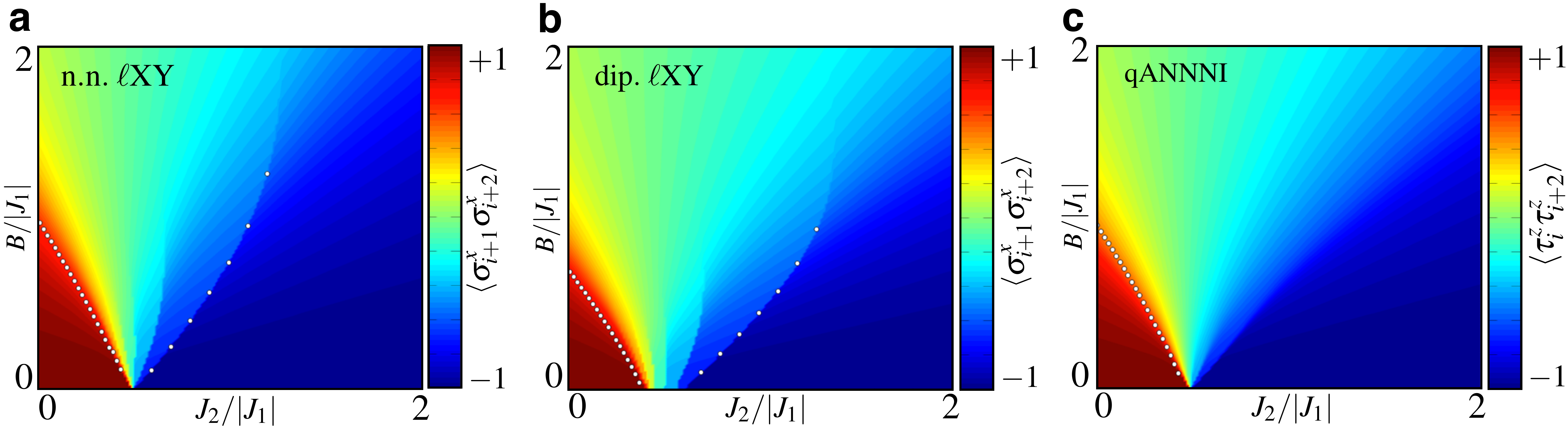}
\caption{\label{CZ} {\bf Two-point correlation functions:} Correlation functions $\langle
\sigma_{i+1}^x \sigma_{i+2}^x\rangle$ in the nearest-neighbor
$\ell$XY model {\bf (a)} and the dipolar $\ell$XY model  {\bf
(b)}, both for $N=15$ and open boundary conditions (averaging
correlations over all possible indices $i$). The white dots mark
the FM-PM transition, obtained by finite-size scaling of the mass
gap, and the AP-FP transition determined via overlap
considerations. {\bf (c)} Correlation functions $\langle \tau_i^z
\tau_{i+2}^z \rangle$ in the dual qANNNI model, for $N=16$ spins
and open boundary conditions. The FM-PM transition line obtained
via finite-size scaling of the mass gap is given by the white
dots. }
\end{figure*}

We now explore numerically the phase diagram of the dipolar XY
model in a longitudinal field~\eqref{dipolar_longitudinal_XY},
assessing how far  it resembles the  qANNNI model~\eqref{qANNNI}.
Let us thus review the features of the qANNNI phase diagram, which
are in one-to-one correspondence with the 2D classical ANNNI
model~\cite{commensurate_incomm_annni_review}.

Setting $B=0$, the qANNNI  Hamiltonian becomes classical, and the
phase diagram is derived easily: For $J_2/|J_1|<0.5$, the system
is in a ferromagnetic (FM) phase, with a two-fold degenerate
ground state having all spins polarized along $z$. For
$J_2/|J_1|>0.5$, the system is the the anti-phase (AP) with a
4-spin unit cell $\uparrow \uparrow \downarrow \downarrow$. For
commensurability purposes, one should take a system with a number
of spins $N$ divisible by 4. For open boundaries, one  obtains a
two-fold degenerate ground state ($\mathcal{Z}_2$ spin-inversion
symmetry), while in the case of periodic boundaries, translational
symmetry yields a four-fold degenerate ground state
($\mathcal{Z}_2\times\mathcal{Z}_2$ symmetry by the Cartesian
product of the spin-inversion symmetries at even and odd lattice
sites). The critical point between the FM phase and the AP
corresponds to the frustration point $J_2/|J_1|=0.5$ addressed in
the previous section, where the model yields an
exponentially-large ground state manifold given
by~\eqref{deg_dual}. If a strong magnetic field $B\gg |J_1|,J_2$
is present, the system will exhibit a paramagnetic (PM) phase,
with two-point spin correlations that decay exponentially fast
with the distance.

 Several studies, including exact diagonalization of an effective domain-wall
 description \cite{rieger96}, finite-size scaling \cite{florencio02}, or recent matrix-product state
 (MPS) calculations \cite{nagy11} predict the direct vicinity of the PM and FM phase, with a second-order quantum phase transition
 along a critical line that bounds the entire FM phase. The FM-PM transition belongs to the same universality class as the
 standard  quantum Ising model~\cite{ising_transverse_field}. As shown by the density-matrix renormalization group (DMRG)
 study of Ref. \cite{beccaria06}, a so-called disorder line, coinciding with the Peschel-Emery line~\cite{peschel_emery_line},
 divides the PM phase into two regimes: In direct vicinity to the FM phase, the PM phase is unmodulated, while for larger values of
 $B$ the exponential decay of correlations is accompanied by a modulation of the correlation function.

For $J_2/|J_1|>0.5$, a so-called floating phase (FP) separates the
AP and the PM phase, as evidenced by the quantum Monte Carlo study
in Ref. \cite{arizmendi91}, by the DMRG studies in Ref.
\cite{nagy11,beccaria07}, or by perturbation theory calculations
in Ref. \cite{dasgupta07}. This intermediate phase is
characterized by modulated, algebraically decaying spin-spin
correlations. The MPS data \cite{nagy11} suggest that the  AP-FP
commensurate-incommensurate Pokrovsky-Talapov transition is
second-order, while the FP-PM transition is believed to be of the
Kosterlitz-Thouless type~\cite{nagy11,beccaria07}. We note that
the determination of the phase diagram in this regime is
difficult, as important quantities like the energy gap strongly
depend on the system size.

Once the properties of the qANNNI model have been discussed, we
can start addressing the effects of the long-range tail through
the numerical study of the  $\ell$XY
model~\eqref{dipolar_longitudinal_XY}.

\subsubsection {Parameter region $J_2/|J_1|<0.5$}

\begin{figure*}
\centering
\includegraphics[width=1\textwidth, angle=0]{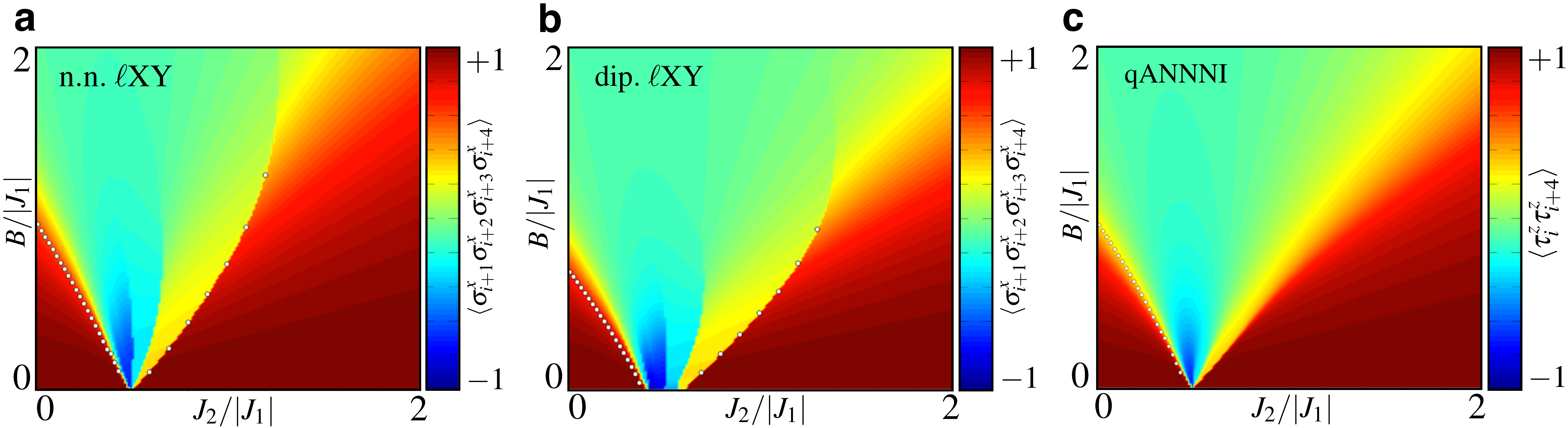}
\caption{\label{CZZ}  {\bf Four-point correlation functions:} Same as in Fig. \ref{CZ}, but
for 4-point correlation functions $\langle \sigma_{i+1}^x
\sigma_{i+2}^x \sigma_{i+3}^x \sigma_{i+4}^x\rangle$ in the
nearest-neighbor $\ell$XY model {\bf (a)} and the dipolar $\ell$XY
model  {\bf (b)}, and 2-point correlation functions $\langle
\tau_i^z \tau_{i+4}^z \rangle$ in the qANNNI model  {\bf (c)}. }
\end{figure*}

In the regime of low frustration, $J_2/|J_1|<0.5$, the dual qANNNI
model behaves similarly to the standard quantum Ising model, with
a  FM-PM second-order transition. Boundary and/or finite-size
effects play a quantitative, but not a qualitative role. The
transition between FM-PM has a relatively sharp signature in the
spin-spin correlation function at sufficiently large distance. As
the PM phase is characterized by exponentially fast correlations,
a drop of the correlations marks the boundary. In the original
model, a correlation $\langle \tau_i^z \tau_{i+\ell}^z \rangle$
translates into a non-local $\ell$-point correlation function
$\langle \sigma_{i+1}^{x} \sigma_{i+2}^{x} \dots
\sigma_{i+\ell}^x\rangle$, which as discussed around
Eq.~\eqref{corr_functions} can be measured in trapped-ion
experiments. We have evaluated these correlations for $\ell=2$ and
$\ell=4$ numerically, and the results are shown in Figs.~\ref{CZ}
and~\ref{CZZ}. In these figures, we compare the results for {\bf
(a)} the nearest-neighbor limit of the $\ell$XY model, {\bf (b)}
its full dipolar version~\eqref{dipolar_longitudinal_XY},  and
{\bf (c)} the corresponding correlations $\langle \tau_i^z
\tau_{i+\ell}^z \rangle$ of the associated dual
model~\eqref{qANNNI}. In all cases, we have applied open boundary
conditions, and averaged the spin-spin correlations over all
lattice sites.  One can see a very nice agreement of these
numerical results between the $\ell$XY model and its qANNNI dual. Note
that, in the case of open boundaries, the qANNNI model of $N$
spins corresponds to the dual of a $\ell$XY  model of $N-1$ spins,
and this is the reason why the spin numbers differ.

As expected, the nearest-neighbor $\ell$XY model and the qANNNI
model behave identically. A comparison of them with the dipolar
$\ell$XY model shows some small differences. For instance, the FM
phase is less extended in the presence of dipolar interactions.
This is due to the fact that in Eq.
(\ref{dipolar_longitudinal_XY}) the dual FM order is established
by the magnetic field term~\eqref{dual_param}, while PM order is
due to interactions. Therefore, the dipolar tail, enhancing
the interaction, acts in favor of the PM phase and shifts the FM-PM
transition towards smaller values of $B/|J_1|$. A similar
reasoning explains why, along $B=0$, the FM phase vanishes for
$J_2/|J_1|<0.5$ in the dipolar model. We also notice that the
critical ``point'' is broadened by dipolar interactions, with a
critical region between the FM-AP transition at $B=0$.

As a possible approach to determine the FM-PM transition line, we
use finite size scaling (FSS) of the mass gap. Therefore, we
define the scaled energy gap
\begin{align}
 \Delta_N(B) = N^z \left(E_1(B)-E_0(B) \right),
\end{align}
with $z$ the dynamical critical exponent, in the following taken
to be 1. The energies $E_1$ and $E_0$ correspond to the first
excited state, and to the ground state of the original $\ell$XY
model~\eqref{dipolar_longitudinal_XY}, respectively. At
criticality, $\Delta_N(B)$ should be independent from the size of
the system, and thus all the curves $\Delta_N(B)$ for different
values $N$ must cross at  one common point. This allows us to
extract the critical field strength $B_{\rm c}$ for a given value
of $J_2$. The data obtained in this way, taking into account
system sizes up to $N=15$, is given by the dots in Figs. \ref{CZ}
and \ref{CZZ}. We highlight how the drop of the four-point
correlator in Fig. \ref{CZZ} agrees extremely  well with the phase
boundary obtained via scaling behavior of the mass gap. Let us
note that the required energy gap energies can be measured
experimentally using the spectroscopic trapped-ion methods put
forth in~\cite{xy_ions_spect,spectroscopy_many_body}. Combined
with the unique property of controlling exactly the number of
spins, would allow the trapped-ion setups to perform the first FSS
in a real experiment.

\begin{figure} [b]
\centering
\includegraphics[width=0.75\columnwidth, angle=0]{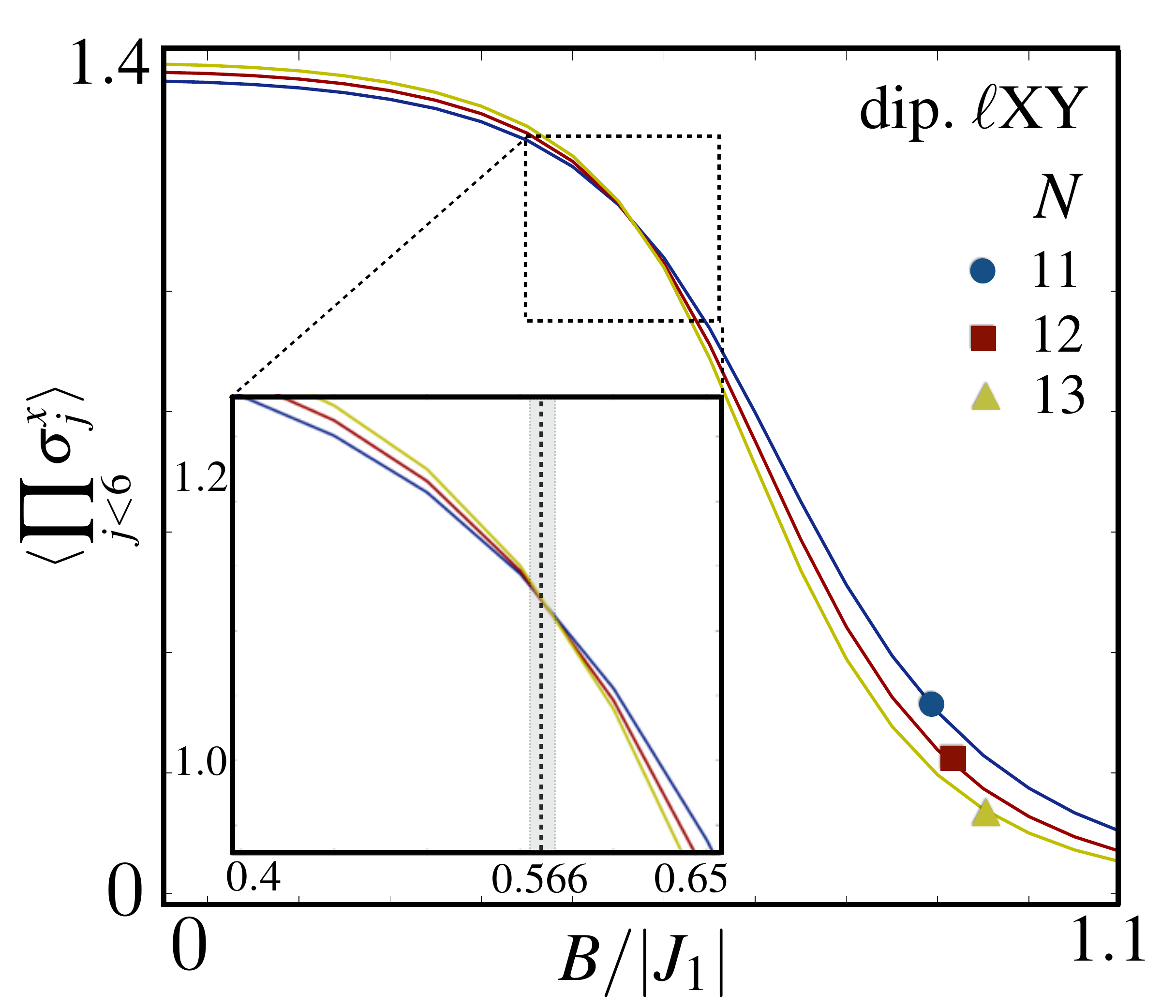}
\caption{\label{fss_corr}  {\bf Finite-size scaling of the non-local correlators:} Finite-size scaling of
the correlation function $N^{1/8} \langle \prod_{i=1}^6 \sigma^z_i
\rangle$, corresponding to the magnetization of a central spin in
the dual model, plotted versus the field strength $B/|J_1|$ at
fixed $J_2/|J_1|=0.1$. The critical field strength $B_{\rm
crit}/|J_1|=0.566$, common to curves corresponding to different
sizes $N$, agrees with the value obtained via finite-size scaling
of the energy gap. The dashed line mark an error interval of width
$\delta B_{\rm crit}/|J_1|=0.01$. }
\end{figure}

As an alternative to the energy gap, other observables, such as
magnetization and correlations, are equally well suited for
performing a FSS, and might be easier to measure in a trapped-ion
experiment. Assuming the $\beta=1/8$ exponent of the quantum Ising
model, one could scale local magnetization of the ground state as
$\langle \tau_i^z\rangle N^{1/8}$, which should become independent
from $N$ at the phase boundary. In the case of the $\ell$XY model,
we would need to examine the dual correlator to  $\langle
\tau_i^z\rangle$, that is, $\langle\prod_{j\leq
i}\sigma_j^x\rangle$. As an example, we have chosen a fixed value
$J_2 = - 0.1 J_1$ in Fig. \ref{fss_corr}, and varying the field
strength $B$, we determine its critical value $B_{\rm c}$ by the
cut of the curves correspoding to different system sizes. The
obtained value, $B_{\rm c}=0.566(10)$, agrees with the
corresponding value obtained via finite-size scaling of the gap,
$B_{\rm c}=0.574(5)$.

FSS also allows to determine critical exponents by assuming a
certain scaling relation \cite{sen2000}, namely
\begin{align}
\label{scaling}
 \Delta_N(B) \sim f(N^{1/\nu}(B-B_{\rm c})),
\end{align}
with $\nu$ the critical exponent of the correlation length,
$B_{\rm c}$ the critical field strength of the FM-PM transition
already calculated, and $f(x)$ a universal scaling function. From
Eq. (\ref{scaling}), it follows that
\begin{align}
\label{nu}
\nu = \frac{\log (N/N')}{\log (\partial \Delta_N/ \partial \Delta_{N'})},
\end{align}
where $\partial \Delta_N \equiv (\partial \Delta_N(B)/ \partial
B)|_{B=B_{\rm c}}$. Using Eq. (\ref{nu}), we have determined $\nu$
from the data $\Delta_N$ with $N$ up to 15, using the $B_{\rm c}$
obtained before. The results are shown in Table \ref{nutab}, for
different values of $J_2$ and evaluated using $N=15$ and $N'=14$,
with errors assuming an uncertainty of $\delta B_{\rm c} = 0.005$.
For all models (qANNNI, nearest-neighbor $\ell$XY, and dipolar
$\ell$XY), we obtain a critical exponent $\nu$ in the range
between 0.9 and 1.1, in accordance with the expected Ising
universality class $\nu=1$. Error estimates suggest that the most
accurate results are obtained, as could be expected, far off the
frustration point $J_2/|J_1|=0.5$, with $\delta \nu=0.02$. With
this, the expected Ising exponent, $\nu=1$, is located within an
interval of up to five times the standard error. We expect that a
systematic underestimation of the critical field strength due to
the finite size of our samples is responsible for this
discrepancy. The best agreement with the Ising value is obtained
in the nearest-neighbor $\ell$XY, with values spreading between
0.94 and 1.01. Remarkably, the dipolar version of the $\ell$XY
model produces results which are slightly better than those
obtained for the original qANNNI model, see Table \ref{nutab}.

\begin{table*}[t]
\begin{tabular}{l|c|c|c|c|c|c}
 $J_2/|J_1|$ & $B_{\rm c}/|J_1|$ & $\nu$           & $B_{\rm c}/|J_1|$ & $\nu$            & $B_{\rm c}/|J_1|$ & $\nu$ \\
         & (n.n. $\ell$XY model)         & (n.n. $\ell$XY model)    & (dipolar $\ell$XY model)      & (dipolar $\ell$XY model)  & (qANNNI model)        & (qANNNI model) \\
 \hline
 1/50  & 0.973(5) & 0.94(2) & 0.695(5) & 0.90(2) & 0.959(5) & 1.09(2) \\
 6/50  & 0.790(5) & 0.95(2) & 0.541(5) & 0.91(2) & 0.780(5) & 1.09(3) \\
 11/50 & 0.595(5) & 0.96(2) & 0.374(5) & 0.92(3) & 0.592(5) & 1.07(4) \\
 16/50 & 0.388(5) & 1.01(6) & 0.184(5) & 0.94(5) & 0.380(5) & 1.13(17)
\end{tabular}
\caption{\label{nutab} Critical field strength $B_{\rm c}$ and critical exponent $\nu$ for the FM-PM transition.}
\end{table*}

\vspace{0.5ex}

\subsubsection{Parameter region  $J_2/|J_1|>0.5$}

For various reasons, the regime $J_2/|J_1|>0.5$ is more difficult
to capture accurately by studying a small-sized system. First of
all, one must be very careful with the number of spins studied,
and the choice of the boundary. Since the anti-phase has a 4-spin
unit cell, one would take system sizes  divisible by 4, such that
the classical ANNNI model ($B=0$) shows the expected FM-AP
transition for both open and periodic boundary conditions at
$J_2/|J_1|=0.5$. However, when we turn to the $\ell$XY model,  the
situation is different: If we again choose $N$ divisible by 4,
only the system with periodic boundary conditions exhibits a
direct FM-AP transition. In contrast, the system with open
boundaries, which is the experimentally-relevant case for the
small-scale trapped-ion magnets, shows an intermediate phase,
extending from $0.5 < J_2/|J_1| < 1$.

The occurrence of this intermediate phase can be understood from
simple energy considerations in the classical system, i.e. in the
$\ell$XY model without YY interactions. The ferromagnetic case is
characterized by a single, fully polarized ground state, whereas
the antiphase has N\'eel order. We assume an intermediate state
with ferromagnetic order for $N_{\rm f}$ subsequent spins and
N\'eel order for $N_{\rm a}=N-N_{\rm f}$ spins. The energy of this
configuration is given by
\begin{align}
 \frac{E}{|J_1|} = N_{\rm a} \left(1- 2 \frac{J_2}{|J_1|} \right) +N \left( \frac{J_2}{|J_1|} -1 \right) - \frac{J_2}{|J_1|},
\end{align}
where we have assumed $N_{\rm a}<N$. For $J_2/|J_1|<0.5$, the
$N_{\rm a}$ term contributes positively to the energy, and thus
the ground state configuration must have $N_{\rm a}=0$, that is,
the system is in the FM phase. For $J_2/|J_1|>0.5$, $N_{\rm a}$
should attain its maximal value, which in the formula above is
$N-2$. The energy for $N_{\rm a}=N$ is simply given by
$E/|J_1|=-(J_2/|J_1|)(N-1)$. Comparison of the two values shows
that the antiphase (i.e. $N_{\rm a}=N$) becomes favorable only for
$J_2/|J_1|>1$, while for $0.5<J_2/|J_1|<1$, the system has two
domains, a ferromagnetic one and a domain in the antiphase.
Independent from the size of the system, the ferromagnetic domain
for $0.5<J_2/|J_1|<1$ is of size $N_{\rm f}=2$, thus for large
systems the intermediate phase becomes indistinguishable from the
antiphase.

\begin{figure*}[t]
\centering
\includegraphics[width=0.75\textwidth, angle=0]{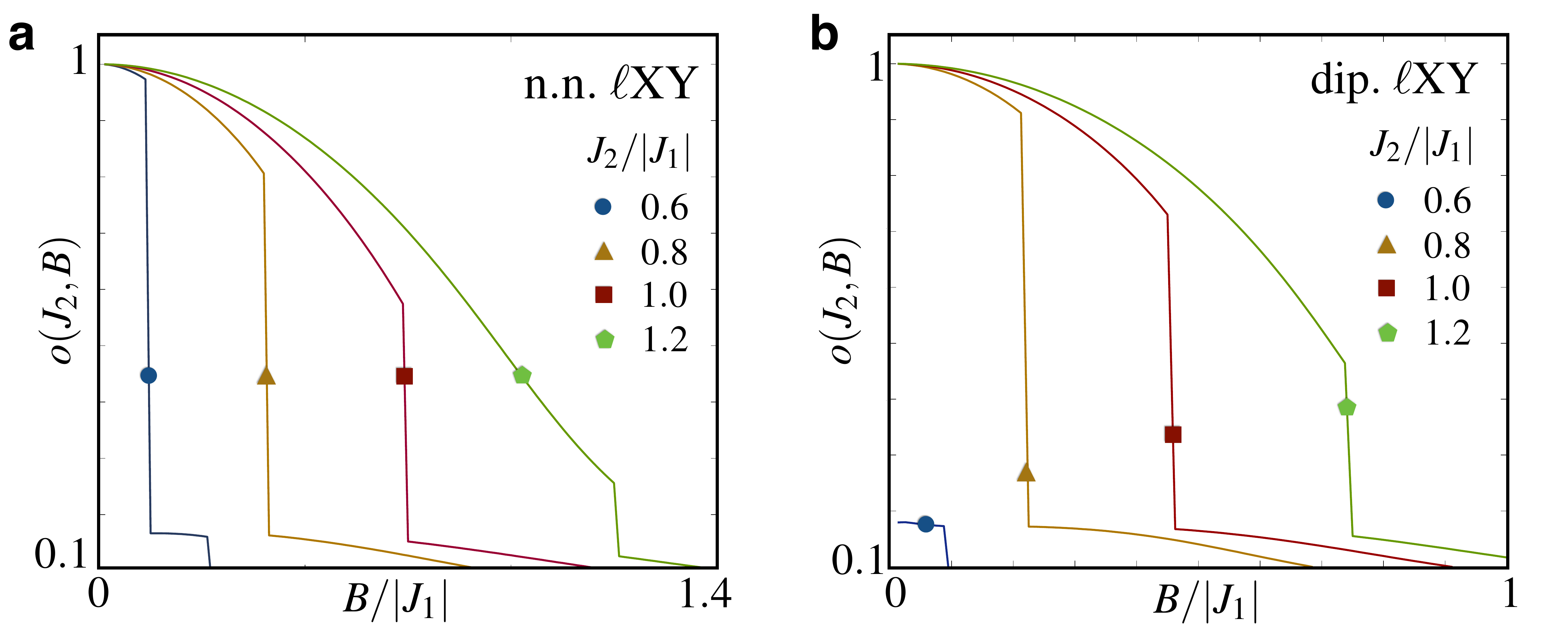}
\caption{\label{ov} {\bf Correlation overlap:} For $N=15$, we plot the
correlation overlap (defined in Eq. (\ref{o})) of the nearest-neighbor (left) and the dipolar (right) $\ell$XY  models. This allows to determine
the phase boundary of the AP for sufficiently small $J_2$. At
larger $J_2$, the vanishing of AP correlations occurs smoothly.}
\end{figure*}

As we announced before, for a system with periodic boundary conditions, the classical energy is always given by
\begin{align}
\frac{ E}{|J_1|} = N_{\rm a} \left(1- 2 \frac{J_2}{|J_1|} \right) +N \left( \frac{J_2}{|J_1|} -1 \right),
\end{align}
for all possible values of $N_{\rm a}$. Therefore, the transition
from a fully ferromagnetic system to a fully anti-phase system
occurs without any intermediate phase.

To solve this disagreement between qANNNI model and $\ell$XY
model, we have to consider  the fact that the dual operators are
bond operators that reside on the links of the
lattice~\eqref{dual_lattice}. Therefore, the open qANNNI chain of
$N=4m$ spins ($m \in \mathbf{Z}^+$) would be the dual model of a
Hamiltonian with $4m-1$ spin operators. Accordingly, we should
find a much better  agreement between the diagram for $N=15$
original spins, and $N=16$ spins in the dual qANNNI model. This
expectation is clearly confirmed by our numerical results (see
Figs. \ref{CZ}, \ref{CZZ}), which show that the odd number
$\ell$XY model does not present an intermediate phase, but simply
the expected FM-AP transition from the dual qANNNI model.

In the quantum case, $B\neq 0$, the system is expected to undergo
two quantum phase transitions: the AP-FP and the FP-PM
transitions. According to an MPS study of the qANNNI model
\cite{nagy11}, the AP-FP transition is of second order, while the
FP-PM is of higher order. Determining the phase boundaries in the
thermodynamic limit is difficult, as due to the modulated nature
of the FP, the minima of the mass gap strongly depend on the
system size $N$. In principle, as was done in
Ref.~\cite{beccaria06} in the framework of a DMRG study for
hundreds of spins, it is still possible to perform finite-size
scaling, if one interpolates between different minima of the mass
gap $\Delta_N(B)$.

Without this costly procedure for extracting the thermodynamic
limit, the AP-FP transition at a fixed finite system size can
easily be detected by looking at the spin-spin correlators:
Sufficiently close to the frustration point, a sudden drop of the
4-point correlations clearly marks a phase boundary in the qANNNI
model, the nearest-neighbor $\ell$XY model, and the dipolar
$\ell$XY model [cf. Fig.~\ref{CZZ}]. To demonstrate that this drop
indeed corresponds to the AP-FP transition, we first note that,
in the language of the $\ell$XY model, the antiphase correlations are
characterised by a N\'eel order of the original spins. Accordingly,
the AP spin-spin correlation function is given by $c_{\rm AP}(d)
:= \langle \sigma_1^x\sigma_{1+d}^x \rangle = (-1)^d$. We thus
introduce the overlap of correlations as
\begin{align}
\label{o}
o(J_2,B) = \sum_{d=1}^{N-1} (-1)^d \langle {\rm GS}(J_2,B) | \sigma_1^x\sigma_{1+d}^x | {\rm GS}(J_2,B) \rangle.
\end{align}
For both the nearest-neighbor and the dipolar model with $N=15$,
the overlap is plotted in Fig.~\ref{ov} as a function of $B$ for
selected values of $J_2$. The drop in overlap corresponds to the
vanishing of AP order, and completely agrees with the drop of
correlations in Figs. \ref{CZ} and \ref{CZZ}. However, for larger
values of $J_2/|J_1|\gtrsim 1$, the function $o(J_2,B)$ becomes
smooth over the full range of $B$, rendering it impossible to
extract critical values of $B$. The presence of dipolar
interactions shifts the AP-FP transition to smaller values of
$-B/|J_1|$, in comparison to the nearest-neighbor $\ell$XY model.

While identifying the anti-phase, and determining its boundary, is
possible even for small systems due to the very characteristic
correlations of the anti-phase, our method does not provide
information about the universality class of the transition. As
done in Ref. \cite{beccaria07,nagy11}, this information can be
obtained from the scaling of the entanglement entropy in
sufficiently large systems. Such procedure also allows to
determine the phase diagram beyond the anti-phase, that is, the
position of the higher-order phase transition between floating phase
and paramagnetic phase.

Although trapped ions yet do not reach the required system size
for identifying the critical behavior at the FP-PM transition, let
us nevertheless provide some numerical evidence by studying the
scaling of the entanglement entropy in the $\ell$XY model.
Therefore, we first define the entanglement entropy through the
$n$th bond of a system with $N$ spins as
\begin{align}
 \label{entent}
 S(N,n) = - {\rm Tr}_{[1,\dots,n]} \left( \rho_{[1,\dots,n]} \ln \rho_{[1,\dots,n]}  \right),
\end{align}
where $\rho_{[1,\dots,n]} \equiv {\rm Tr}_{[n+1,\dots,N]}
\ket{\Psi}\bra{\Psi}$ the density matrix of the left part of the
system (containing spins 1 to $n$). The system is assumed to be in
a pure state  $\ket{\Psi}$. Quantum criticality is reflected by a
scaling of the entanglement entropy as
\begin{align}
 \label{critscaling}
 S(N) \simeq \frac{c}{6}\log_2(N) + {\rm const.},
\end{align}
with $c$ being the central charge of the underlying conformal
field theory, and the additive constant being non-universal and
depending on $n$. For non-critical systems the correlation length
$\xi$ is finite, and finite-size corrections described by a
universal function $s(N/\xi)$ need to be taken into account.
Following Ref. \cite{beccaria07}, we assume a scaling behavior
\begin{align}
 \label{noncritscaling}
 S(N) \simeq \frac{c}{6}\log_2(N) - \frac{c}{6}\ln\left (\frac{N}{\xi}+ e^{-\alpha N/\xi} \right)  + {\rm const.},
\end{align}
where the parameter $\alpha$ will be taken as 1. We will also fix
$n$ at the center of the chain, that is $n=(N+1)/2$, as we will
focus on odd $N$. Eq. (\ref{noncritscaling}) implies that $S(N)$
saturates for finite $\xi$ at sufficiently large $N$, while Eq.
(\ref{critscaling}) implies a logarithmic divergence for critical
systems.

Using the DMRG code implemented in the iTensor library
\cite{itensor}, we have studied the $\ell$XY model for system
sizes up to $N=327$, with up to 200 states kept in a DMRG sweep.
Instead of studying the full dipolar model, we have cut the
long-range tail of the interactions, taking into account only
nearest- and next-to-nearest interactions. For a fixed parameter
$J_2/|J_1|$, we calculate $S(N)$ for different magnetic field
strengths $B$. From the behavior of $S(N)$, we are able to
determine the critical regions: Up to a first critical field
strength $B_{\rm c,1}$, the entanglement entropy takes a
relatively small value and does not scale with $N$. This behavior
indicates a vanishing correlation length, and agrees with the
structure of the anti-phase. For slightly larger values of $B$,
the entanglement entropy still remains constant up to some value
$N$, suggesting that sufficiently small systems still exhibit
anti-phase behavior, while larger systems already have floating
phase FP behavior. For example, for $J_2/|J_1|=1$, we get $B_{\rm
c,1}\approx 0.25|J_1|$, but anti-phase behavior persists in
systems of size $N=55$ for field strengths up to $B=0.26|J_1|$, in
systems of size $N=31$ up to $B=0.27|J_1|$, and in systems as
small as $N=11$ even up to $B=0.5|J_1|$.

\begin{figure}[t]
\centering
\includegraphics[width=0.8\columnwidth, angle=0]{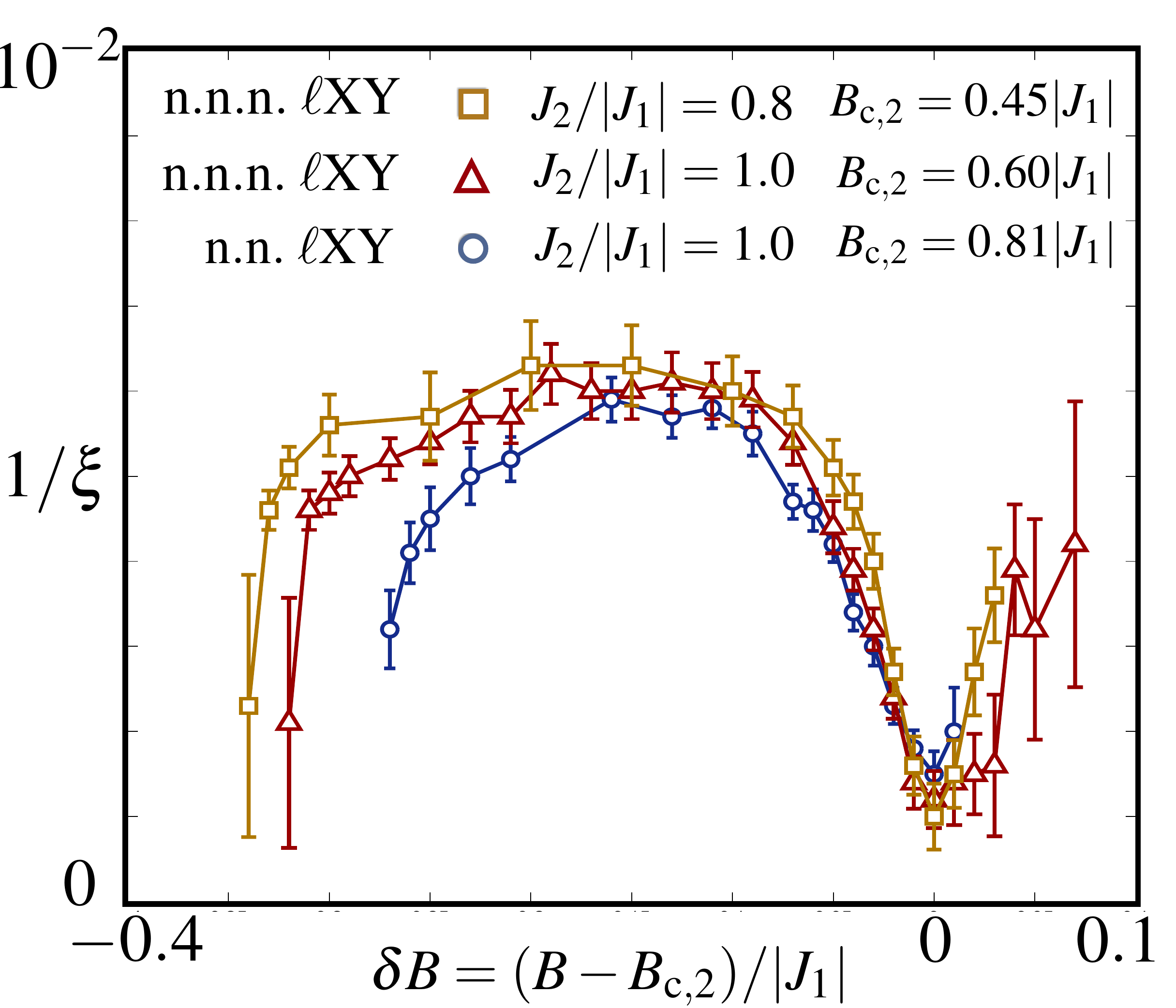}
\caption{\label{xi_fig} {\bf DMRG  calculations of inverse correlation lengths:} We plot the inverse of the
correlation length, $1/\xi$, obtained via scaling of the
entanglement entropy, for different models and different values
$J_2$. ``NN+NNN'' refers to the $\ell$XY model with dipolar
interactions truncated at the order of next-nearest neighbors.
``NN only'' refers to the $\ell$XY model with purely nearest
neighbor interactions. We plot $1/\xi$ as a function of $\delta B
\equiv (B-B_{\rm c,2})/|J_1|$, where $B_{\rm c,2}$ is determined
as the field strength with maximum correlation length.}
\end{figure}

In the floating point regime, the scaling of $S(N)$ is well
described by Eq. (\ref{noncritscaling}), compatible with the
expected central charge $c=1$. Fitting our data to Eq.
(\ref{noncritscaling}) yields a finite correlation length, plotted
in Fig. \ref{xi_fig}. When $B$ approaches a second critical value
$B_{\rm c,2}$, the correlation length significantly increases,
indicating quantum criticality in the vicinity of the FP-PM
transition. For even larger values of $B>B_{\rm c,2}$, $S(N)$
again becomes constant, though now in the limit of large $N$. This
behavior is expected for a sufficiently large paramagnetic system,
characterized by a small but finite correlation length.

Qualitatively, the described behavior is found for different
$J_2/|J_1|$. We have focussed on two values $J_2/|J_1|=0.8$ and
$J_2/|J_1|=1$. In the former case, we obtain $B_{\rm
c,1}=0.10|J_1|$ and $B_{\rm c,2}=0.45|J_1|$. In the latter case,
we obtain $B_{\rm c,1}=0.25|J_1|$ and $B_{\rm c,2}=0.60|J_1|$.
Remarkably, the extent of the floating phase along $B$, that is
$B_{\rm c,2}-B_{\rm c,1}$, is the same for both parameters $J_2$,
in agreement with earlier findings for the qANNNI model
\cite{beccaria07}.

Finally let us turn our attention to the nearest-neighbor $\ell$XY
model, which we have studied for $J_2/|J_1|=1$. We find the same
behavior as before for the system with nearest and next-to-nearest
interactions, but the critical field strengths $B_{\rm c,1}$ and
$B_{\rm c,2}$ are shifted towards larger values (0.52 and 0.81 for
$J_2/|J_1|=1$). The extent of the floating phase is slightly
smaller than in the system with nearest-neighbor and
next-nearest-neighbor interactions.

As a summary, our results suggest that, sufficiently far from the
frustration point, the  dipolar  $\ell$XY model has no qualitative
changes with respect to the qANNNI phase diagram. We find clear
evidence for the paramagnetic to ferromagnetic phase transition,
as well as for the transition from the anti-phase to the floating
phase, even in small-sized systems. DMRG calculations for larger
systems demonstrate that also the  phase transition from the
floating phase to the paramagnetic phase persists in the presence
of interactions beyond nearest neighbors.

\section{Dual quantum simulators of multi-spin interactions}
\label{sec:qs_multi_spin}

In this section, we want to show that the dual quantum
simulator~\eqref{dual_lXY} can be used to explore  the phase
diagram of quantum Ising models with multi-spin interactions.

In order to achieve the desired Hamiltonian, we shall focus on a
particular trapped-ion realization of the above scheme that uses
two different hyperfine levels within the atomic ground state
manifold to encode the pseudo-spin. In this case,  the
spin-dependent dipole forces~\eqref{spin_dep_forces} that couple
to both radial phonon branches,  $F_{i,n}^{\alpha}$ with
$\alpha\in\{x,y\}$, arise from a combination of two-photon Raman
transitions, each of which requires a pair of laser beams tuned
far from the resonance of a dipole-allowed transition to an
excited state~\cite{wineland_review}. If the effective wave vector
of the interfering laser beams,  $\Delta
\boldsymbol{k}^\alpha=\boldsymbol{k}^\alpha_{\rm
L,1}-\boldsymbol{k}^\alpha_{\rm L,1}$, has a component along the
ion-trap axis, ${\bf e}_z\cdot\Delta \boldsymbol{k}^\alpha=\Delta
k^\alpha_z\neq0$, and the two-photon Raman transitions lie far
away from the axial sidebands, one recovers the desired $\ell$XY
model~\eqref{dipolar_longitudinal_XY}, but with modified spin-spin
interactions~\eqref{dip_decay}, namely \beq
\label{dip_decay_modulated} J_{ij}^{\alpha}\approx
\frac{J_0^\alpha\cos\left(\Delta
k^\alpha_z\big(z_i^0-z_j^0\big)\right)}{|z_i^0-z_j^0|^3},\hspace{2ex}
J_0^\alpha>0,\hspace{2ex}\alpha\in\{x,y\}. \eeq We thus obtain
periodically-modulated spin-spin interactions that still decay
with a dipolar power law. By controlling the ion lattice spacing,
or the direction of the laser beams, one can set  $\Delta
k^x_z\big(z_i^0-z_{i+1}^0\big)=\frac{\pi(2m+1)}{2}$ for some
integer $m\in\mathbb{Z}^+$, such that the nearest-neighbor
spin-spin couplings are inhibited $J_{i,i+1}^{x}\approx0$. The
fulfilment of this condition requires an homogeneous lattice
spacing in the ion crystal, which can be achieved by using
micro-fabricated surface traps~\cite{surface_traps} or by
segmenting the axial electrodes in linear Paul
traps~\cite{homogeneous_lattice_ions}.

Under these conditions, we can rearrange  the dual
Hamiltonian~\eqref{dual_lXY} as follows $H^{\rm dual}_{\rm \ell
XY}=H_{\rm qIMS}(-h,J_{i,i+2}^x,J_{i,i+1}^y)+\Delta \tilde{H}$,
where we have introduced the quantum Ising model with multi-spin
interactions (qIMS), namely \beq \label{qIMS} H_{\rm
qIMS}(\tilde{J}_2,\tilde{J}_4,\tilde{B})=\sum_i\left(\tilde{J}_2\tau_i^z\tau_{i+1}^z+\tilde{J}_4\tau_i^z\tau_{i+1}^z\tau_{i+2}^z\tau_{i+3}^z+B\tau_i^x\right).
\eeq Here, the Ising interaction $\tilde{J}_2<0$ is a
ferromagnetic nearest-neighbor  coupling that corresponds to the
longitudinal field  of the original
Hamiltonian~\eqref{dipolar_longitudinal_XY},  $\tilde{J}_4>0$ is
an anti-ferromagnetic 4-spin Ising coupling that corresponds to
the next-nearest-neighbor XX couplings of the original model,  and
$\tilde{B}$ is  a transverse magnetic field  that corresponds to
the nearest-neighbor YY couplings of the original model.  This
parameter equivalence under the duality transformation is \beq
\label{dual_param_qIMS} \tilde{J}_2=-h,\hspace{2ex}
\tilde{J}_4=J_{i,i+2}^x,\hspace{2ex}\tilde{B}=J_{i,i+1}^y. \eeq
 Let us emphasize that the competition of  the Ising interactions, characterized by  $\tilde{J}_4/|\tilde{J}_2|$, corresponds to the ratio
 $J_{i,i+2}^x/h$ in the original model, which can be tuned to any particular value. The same occurs for the ratio $\tilde{B}/|\tilde{J}_2|$,
 which corresponds to  $J_{i,i+1}^y/h$, such that the whole phase diagram of the qIMS can be addressed with the dual quantum simulator.

 In addition to the qIMS Hamiltonian, the dipolar tail of the interactions leads to the following perturbation
\beq \label{pert}
\tilde{H}=\sum_{i=1}^{N+1}\sum_{j=i+3}^{N+1}J_{ij}^x\tau_{i-1}^z\tau_{i}^z\tau_{j-1}^z\tau_{j}^z+\sum_{i=1}^{N+1}\sum_{j=i+2}^{N+1}J_{ij}^yJ_{ij}^y\prod_{i\leq
k< j}\tau_k^x, \eeq where some of the couplings also vanish due to
the periodic modulation~\eqref{dip_decay_modulated}, namely
$J_{ij}^x\approx 0$ for all $j=i+(2m+1)$ with $m\in\mathbb{Z}^+$.
Due to the fast dipolar decay~\eqref{dip_decay_modulated}, one can
argue once more that this perturbation will not modify in any
essential manner the phase diagram of the qIMS model~\eqref{qIMS},
such that the dual quantum simulator~\eqref{dual_lXY} also gives
access to the physics of these exotic quantum magnets.

In the absence of 4-body  couplings $\tilde{J}_4=0$,  this
model~\eqref{qIMS} corresponds to the standard quantum Ising
model~\cite{ising_transverse_field}, displaying a second-order
phase transition at $\tilde{J}_2=\tilde{B}$ in the Ising
universality class~\cite{ising_2d}. In the absence of pair-wise
couplings $\tilde{J}_2=0$, the model corresponds to the
quantum-mechanical version of a  classical Ising model in a square
lattice~\cite{classical_multi_spin} describing stacked Ising
chains with  4-body couplings at finite temperatures, as can be
proved by applying the quantum-classical
mapping~\cite{quantum_classical_ising} in the present
case~\cite{q_multi_spin_Ising,q_multi_spin_Ising_duality}. The
model has a $\mathcal{Z}_2\times\mathcal{Z}_2\times\mathcal{Z}_2$
symmetry (i.e. cartesian product of  spin inversions for the 3
different spin pairs in each 4-site partition of the lattice),
which leads to an eight-fold degenerate ground state for
$\tilde{B}=0$, and a first-order phase transition at
$\tilde{J}_4=\tilde{B}$ in the universality class of the $q=8$
Potts model~\cite{first_order_mc}. As occurred for the qANNNI
model~\eqref{qANNNI}, one thus expects that the competition of
both Ising interactions, and their interplay with the quantum
fluctuations brought by the transverse field, must lead to  very
interesting critical phenomena. In contrast to the frustrated
Hamiltonian, this quantum multi-spin Ising model has not been
studied in such detail. FSS studies with up to $N=16$ spins
already point towards very interesting critical phenomena: The
nature of the quantum phase transitions as a function of
$\tilde{B}$ changes from first to second order around
$\tilde{J}_4/|\tilde{J}_2|=1/2$~\cite{2_4_Ising}. However, the
limited finite sizes did not allow for accurate studies close to
the interesting point  $\tilde{J}_4/|\tilde{J}_2|=1/2$, which
classically $\tilde{B}=0$ leads to a largely-degenerate ground
state that can give rise to a variety of phases upon switching the
transverse magnetic field. We believe that a more careful analysis
of this region would be very interesting, and we hope that this
manuscript will stimulate future work in that direction.

From the experience gained by the exhaustive numerical study of
the previous section, we conjecture that the quantum simulator of
the quantum Ising model with multi-site interactions~\eqref{qIMS}
will not be compromised by the additional dipolar
tail~\eqref{pert}, such that the above interesting region can be
characterized experimentally using the accessible non-local order
parameters~\eqref{corr_functions}, or the FSS of other
spectroscopic observables, characteristic of current ion-trap
experiments.

\section{Conclusions and Outlook}
\label{sec:conclusions}

In this manuscript, we have presented an alternative route to
build quantum simulators of exotic magnetism by exploiting the
notion of quantum dualities. In certain situations, such as the
ones discussed in this work, such duality transformations allow
one to explore interesting models that involve frustration or
multi-spin interactions, and their interplay with quantum
fluctuations, by focusing on different spin models that are
simpler to implement in a particular experimental platform. This
dual approach has  one important caveat: measurements of highly
non-local observables should be feasible in the particular
experiment. This makes trapped-ion setups ideally suited for this
playground of dual quantum simulation.

We have focused on two particular dual quantum simulators which
can be implemented using state-of-the-art technology in linear ion
crystals, and which allow one to explore paradigmatic models of
frustration: the quantum axial nearest-neighbor Ising model, and
the quantum Ising model with competing 2- and 4-body interactions.
In the former case, we have made a careful numerical study to
prove the validity of our scheme, which takes into account
relevant perturbations that occur naturally in the trapped-ion
scenario.

Although we have focused on these two particular examples for ion
chains, the notion of  dual quantum simulators can be certainly
applied to other models, such as the  quantum spin liquids,
topologically-ordered phases, or Ising lattice gauge theories
mentioned in the introduction. Provided that quantum magnets are
finally synthesized in Penning or surface traps, this
quantum-duality approach will also become relevant for other
lattice geometries, which may define an alternative route to these
exotic quantum many-body phenomena.

Finally, let us mention that in the recent years another platform
for long range spin models has been proposed, namely {\it
ultracold atoms in nanostructures}, or more precisely ultracold
atoms trapped in a vicinity of tapered fibers and optical crystals
(band gap materials). The experimental progress in coupling of
ultracold atomic gases to nanophotonic waveguides  is very rapid
(for a recent review cf. \cite{Atomlight}). The ideas and
proposals concerning realization of long range spin models were
developed for instance in Refs. \cite{Chang, Chang1, Chang2}, and it would be interesting to explore the possibilities of quantum dualities in this context.

\acknowledgements We acknowledge discussions with Andrew Ferris,
Luca Tagliacozzo, and Shi Ju Ran. A.B. acknowledges  support from
Spanish Mineco Project FIS2012-33022, and the CAM Research Network
QUITEMAD+. T.G. and M.L. acknowledge support from EU grants
OSYRIS (ERC-2013-AdG Grant No.  339106), SIQS (FP7-ICT-2011-9 No.
600645),  QUIC (H2020-FETPROACT-2014 No.641122), EQuaM
(FP7/2007-2013 GrantNo. 323714), Spanish Ministry grant FOQUS
(FIS2013-46768-P), Generalitat de Catalunya grant SGR 874, and
Fundaci\'o  Cellex.

\end{document}